%% file: EnumeratingHoloHF.tex
\documentclass[journal=jctcce,layout=twocolumn]{achemso}
\usepackage{natmove}
\usepackage{amsfonts,amsmath,amsbsy,upgreek,color}
\usepackage{textcomp}
\usepackage{dblfloatfix}
\usepackage[]{graphicx}
\usepackage[version=3]{mhchem}

\usepackage{placeins}
\usepackage{multirow,booktabs}
\usepackage{chemfig}
\usepackage{siunitx}

\usepackage{soul}

\makeatletter
\newcommand*\rel@kern[1]{\kern#1\dimexpr\macc@kerna}
\newcommand*\widebar[1]{%
  \begingroup
  \def\mathaccent##1##2{%
    \rel@kern{0.8}%
    \overline{\rel@kern{-0.8}\macc@nucleus\rel@kern{0.2}}%
    \rel@kern{-0.2}%
  }%
  \macc@depth\@ne
  \let\math@bgroup\@empty \let\math@egroup\macc@set@skewchar
  \mathsurround\z@ \frozen@everymath{\mathgroup\macc@group\relax}%
  \macc@set@skewchar\relax
  \let\mathaccentV\macc@nested@a
  \macc@nested@a\relax111{#1}%
  \endgroup
}
\makeatother

\usepackage{caption} 
\usepackage{subcaption}
\usepackage{cleveref}
\usepackage{float} 

\usepackage{mathtools}

\usepackage{tikz}
\usetikzlibrary{backgrounds}      
\usetikzlibrary{shapes.misc}
\usepackage{fancybox,framed}
\newlength{\figurewidth}
\tikzset{cross/.style={cross out, draw=black, minimum size=1*(#1-\pgflinewidth), inner sep=0pt, outer sep=0pt},
cross/.default={0.5cm}} 
\usetikzlibrary{patterns}

\title[Holomorphic HF]{Holomorphic Hartree--Fock Theory: The Nature of Two-Electron Problems}
\author{Hugh~G.~A.~Burton}
\email{hb407@cam.ac.uk}
\affiliation{Department of Chemistry, Lensfield Road, Cambridge, CB2 1EW, UK}
\author{Mark~Gross}
\affiliation{DPMMS, CMS, Wilberforce Road, Cambridge, CB3 0WB, UK}
\author{Alex~J.~W.~Thom}
\affiliation{Department of Chemistry, Lensfield Road, Cambridge, CB2 1EW, UK}
\date{\today}
\begin{document}

\begin{abstract}
We explore the existence and behaviour of holomorphic restricted Hartree--Fock (h-RHF) solutions for two--electron problems.
Through algebraic geometry, the exact number of solutions with $n$ basis functions is rigorously identified as $\frac{1}{2}(3^n - 1)$, proving that states must exist for all molecular geometries.
A detailed study on the h-RHF states of \ce{HZ} (STO-3G) then demonstrates both the conservation of holomorphic solutions as geometry or atomic charges are varied and the emergence of complex h-RHF solutions at coalescence points.
Using catastrophe theory, the nature of these coalescence points is described, highlighting the influence of molecular symmetry.
The h-RHF states of \ce{HHeH^2+} and \ce{HHeH} (STO-3G) are then compared, illustrating the isomorphism between systems with two electrons and two electron holes.
Finally, we explore the h-RHF states of ethene (STO-3G) by considering the $\uppi$ electrons as a two--electron problem, and employ NOCI to identify a crossing of the lowest energy singlet and triplet states at the perpendicular geometry.
\end{abstract}
\maketitle

\section{Introduction}

Hartree--Fock theory is ubiquitous in quantum chemistry.
Representing the many-electron wave function as a single Slater determinant, the Hartree--Fock approximation provides a mean-field description of molecular electronic structure.\cite{SzaboOstlund}
Through the long established self-consistent field method (SCF), the Hartree--Fock energy is minimised with respect to variations of a set of orbitals expressed in a given finite basis set.
This optimal set of orbitals therefore exists as a stationary point of the energy.\cite{Hall1951, Roothaan1951}
However, it is less widely appreciated that the nonlinear form of the SCF equations can lead to convergence onto a range of different solutions.\cite{Lions1987, Thom2008} 
These solutions represent additional local minima, maxima or saddle points of the energy.

Through methods including SCF metadynamics\cite{Thom2008} and the maximum overlap method (MOM)\cite{Gilbert2008}, locating higher energy stationary points has become relatively routine, and several authors have sought to interpret these as physical excited states.\cite{Gilbert2008, Besley2009, Barca2014, Peng2013, Glushkov2015}
However, for many systems there exist multiple low energy solutions that may cross as the geometry changes, presenting a dilemma when correlation techniques require a single reference determinant to be chosen.\cite{Thom2009}
Recently there has been increasing interest in using multiple Hartree--Fock states as a basis for non-orthogonal configuration interaction (NOCI) calculations, providing a more egalitarian treatment of individual low energy SCF solutions.\cite{Malmqvist1986, Ayala1998, Thom2009, Mayhall2014, Sundstrom2014, Jake2017} 

Since each Hartree--Fock state is itself an mean-field optimised solution, excited configurations can be accurately represented in the NOCI basis set\cite{Sundstrom2014}.
Consequently, NOCI also provides an alternative to the multiconfigurational Complete Active Space SCF (CASSCF)\cite{Helgaker} approach
using an ``active space'' of relevant Hartree--Fock determinants.
This results in wave functions that reproduce avoided crossings and conical intersections at a similar scaling to the SCF method itself.\cite{Thom2009, Jake2017} 
Furthermore, the inherent multireference nature of NOCI enables strong static correlation to be captured, whilst additional dynamic correlation can subsequently be computed using the pertubative NOCI-MP2 approach.\cite{Yost2013, Yost2016}

Unless it is strictly enforced, there is no guarantee that SCF solutions possess the same symmetries as the exact wave function.\cite{Fukutome1981, Jimenez-Hoyos2012}
Restricted Hartree--Fock (RHF) wave functions, for example, are eigenfunctions of the spin operator $\hat{S}^2$ but may break the molecular point group symmetry at singlet instabilities\cite{Mestechkin1978,Mestechkin1979,Mestechkin1988}.
In constrast, the unrestricted Hartree--Fock (UHF) approach allows the wave function to break both spatial and $\hat{S}^2$ symmetry, leading to spin contaminated states containing a mixture of singlet and triplet components.\cite{Fukutome1981,Fukutome1974, Fukutome1975}
Alongside capturing static correlation, including symmetry broken SCF states in a NOCI calculation allows spatial symmetry to be restored and reduces spin contamination in a similar style to the Projected\cite{Lowdin1955, Scuseria2011, Ellis2013, Jimenez-Hoyos2012} and Half-Projected\cite{Smeyers1973,Smeyers1974,Cox1976,Smeyers1976} Hartree--Fock approaches.
However, as a projection after variation approach, NOCI retains the size--consistency of the SCF determinants to provide size--consistent approximations for singlet and triplet wave functions.

Crucially, NOCI requires the existence of multiple Hartree--Fock solutions across all molecular geometries of interest to ensure the basis set size is consistent and prevent discontinuous NOCI energies.\cite{Thom2009, Mayhall2014}
There is, however, no guarantee that Hartree--Fock states must exist everywhere, and in fact they often vanish as the geometry is varied.
This is demonstrated by the coalescence of the low energy UHF states with the ground state RHF solution at the Coulson--Fischer point in \ce{H2}\cite{Coulson1949}, although further examples are observed in a wide array of molecular systems.\cite{Dunietz2003, Cui2013, Mori-Sanchez2014}

To construct a continuous basis set of SCF solutions for NOCI, Thom and Head--Gordon proposed that Hartree--Fock states may need to be followed into the complex plane.\cite{Thom2009}
We have recently reported a holomorphic Hartree--Fock theory as a method for constructing a continuous basis of SCF determinants in this manner.\cite{Hiscock2014, Burton2016}
In holomorphic Hartree--Fock theory, the complex conjugation of orbital coefficients is removed from the standard Hartree--Fock energy to yield a complex differentiable function which we believe has a constant number of stationary points across all geometries.\cite{Hiscock2014}
Using a revised SCF method, we have demonstrated the existence of holomorphic UHF (h-UHF) solutions for \ce{H2}, \ce{H4^2+} and \ce{H4}\cite{Burton2016}.
The h-UHF states exist across all geometries, corresponding to real Hartree--Fock solutions when these are present and extending into the complex plane when the real states disappear.

Despite the promise of holomorphic Hartree--Fock theory, there is currently limited understanding about the nature of holomorphic solutions.
For example, underpinning this theory we believe that the number of stationary points of the holomorphic energy function is constant (including solutions with multiplicity greater than one), and thus states must exist for all geometries.

In the current work, we attempt to understand the simplest application of holomorphic Hartree--Fock theory by investigating the holomorphic RHF (h-RHF) solutions to two--electron problems.
First, we outline the key concepts of the theory before providing a derivation for the exact number of h-RHF states for two electrons in $n$ basis functions.
In doing so we demonstrate that this number is constant for all geometries. 
We then study the full set of h-RHF states for \ce{HZ}, \ce{HHeH^2+} and \ce{HHeH} (STO-3G), investigating the behaviour of these states as molecular geometry or atomic charges are varied, and discussing the isomorphism between systems with two electrons and two electron holes.
Finally, we investigate the h-RHF states of ethene (STO-3G) and demonstrate the application of NOCI to its internal rotation by considering the $\uppi$ and $\uppi^*$ orbitals with frozen core and virtual orbitals as a two--electron SCF problem.

\section{Holomorphic Hartree--Fock Theory}
We begin with an orthonormal set of $n$ real single-particle basis functions, denoted $\{ \chi_{\mu} \}$, from which the closed-shell molecular orbitals can be constructed as
\begin{align}
\phi_{i} = \sum_{\mu}^{n} \chi_{\mu} c_{\mu i}.
\label{eq:MolecularOrbitals}
\end{align}
In standard RHF theory, to ensure orthogonality of molecular orbitals, the orbital coefficients $\{ c_{\mu i} \}$ are elements of a unitary matrix
\begin{align}
\sum_{\mu}^{n} c_{\mu i}^* c_{\mu j}^{\vphantom{*}} = \delta_{i j}.
\label{eq:UnitaryConstraint}
\end{align}
The density matrix is then constructed as $P_{\mu \nu} = \sum_{i}^{N} c_{\mu i}^{\vphantom{*}} c_{\nu i}^*$, where $N$ is the number of occupied spatial orbitals, and the Hartree--Fock energy  is given by
\begin{align}
E &= h_{0} + 2  \sum_{\mu \nu}^n P_{\mu \nu} h_{\mu \nu} \nonumber \\
  &+ \sum_{\mu \nu \sigma \tau}^n P_{\mu \nu} \left[ 2( \mu \nu | \sigma \tau) - ( \mu \tau | \sigma \nu) \right] P_{\sigma \tau}.
\label{eq:ConvetionalHartreeFockE}
\end{align}
where $h_{0}$ is the nuclear repulsion, $h_{\mu \nu}$ are the one-electron integrals and $( \mu \nu | \sigma \tau)$ are the two-electron integrals.
RHF solutions then exist as stationary points of Equation \ref{eq:ConvetionalHartreeFockE} constrained by Equation \ref{eq:UnitaryConstraint}. 

Conventionally, the Hartree--Fock energy function is considered only over the domain of real orbital coefficients.
Extending this domain to the complex plane turns Equation \ref{eq:ConvetionalHartreeFockE} into a function of several complex variables  $\{ c_{\mu k} \}$ and their complex conjugates $\{ c_{\mu k}^* \}$.
However, since the dependence on  $\{ c_{\mu k}^* \}$ violates the Cauch--Riemann conditions,\cite{Fischer} the Hartree--Fock energy is usually interpreted as a function of the real variables $\{ \Re \left[ c_{\mu k} \right] \}$ and $\{ \Im \left[ c_{\mu k} \right] \}$ to ensure gradients are well-defined.
Consequently, $E$ remains a polynomial of only real variables and we cannot expect the number of solutions to be constant for all geometries, as previously demonstrated in the single variable case.\cite{Hiscock2014}

In holomorphic Hartree--Fock, states that disappear can be followed into the complex plane by defining a revised complex-analytic energy as a function of the holomorphic density matrix 
$\widetilde{P}_{\mu \nu} = \sum_{i}^{N} c_{\mu i} c_{\nu i}$,
where now the complex conjugation of orbital coefficients has been removed.
Since $\widetilde{P}$ is a complex symmetric matrix, its eigenvectors --- which form the holomorphic one-electron orbitals --- must be complex orthogonal\cite{Craven1969, Gantmacher} and the orbital coefficients are elements of a complex orthogonal matrix such that
\begin{align}
\sum_{\mu}^{n} c_{\mu i} c_{\mu j} = \delta_{i j}.
\label{eq:ComplexOrthogonalConstraint}
\end{align}
The h-RHF energy is then defined in terms of $\widetilde{P}$ as
\begin{align}
\label{eq:HolomorphicHartreeFockE}
 \widetilde{E} &= h_{0} + 2 \sum_{\mu \nu}^n h_{\mu \nu} \widetilde{P}_{\mu \nu} \nonumber \\
 &+ \sum_{\mu \nu \sigma \tau}^n \widetilde{P}_{\mu \nu} \left[2(\mu \nu | \sigma \tau) - (\mu \tau | \sigma \nu) \right]  \widetilde{P}_{\sigma \tau}.
\end{align}
With no dependence on the complex conjugate of orbital coefficients, this function is a complex analytic polynomial which, by taking inspiration from the fundamental theorem of algebra, we believe must have a constant number of solutions at all geometries.\cite{Hiscock2014, Burton2016}

\section{Enumerating the h-RHF states}

The closed-shell h-RHF approach with two-electrons, described by a single molecular orbital $\phi$ and $n$ orbital coefficients $\{ c_{\mu} \}$, provides the simplest system in which we can consider proving the number of holomorphic Hartree--Fock states is constant for all geometries.
In this case, $\phi$ is constructed from a linear combination of $n$ real orthogonal basis functions as
\begin{equation}
\phi\left( \mathbf{r} \right) = \sum_{\mu=1}^n  c_{\mu} \chi_{\mu} \left( \mathbf{r} \right),
\label{eq:basis_expansion}
\end{equation}
with the requirement for complex orthonormalization introducing the constraint
\begin{align}
\sum_{\mu=1}^n c_{\mu}^2 = 1 .
\label{eq:normalisation}
\end{align}
The holomorphic restricted Hartree--Fock energy is given by the polynomial
\begin{align}
\label{eq:HoloEnergy}
\widetilde{E} \left( c_1,\dots, c_n \right) &= h_{0} + 2 \sum_{\mu, \nu=1}^n h_{\mu \nu} c_{\mu} c_{\nu} \nonumber \\
&+ \sum_{\mu, \nu, \sigma, \tau=1}^n h_{\mu \nu \sigma \tau} c_{\mu} c_{\nu} c_{\sigma} c_{\tau} ,
\end{align}
where $ h_{\mu \nu \sigma \tau} = 2(\mu \nu | \sigma \tau) - (\mu \tau | \sigma \nu)$, and the h-RHF states exist as stationary points constrained by Equation \ref{eq:normalisation}.
Identifying the number of these stationary points can be achieved through the mathematical framework of algebraic geometry.\cite{Hartshorne}

Algebraic geometry forms a vast and complex field, encompassing the study of solutions to systems of polynomial equations in an affine or projective space.
Affine spaces provide a generalisation to Euclidean space independent of a specific coordinate system. 
An $n$-dimensional affine space $\mathbb{A}^n = \mathbb{C}^n$ is described by the $n$-tuples $(a_1, \dots, a_n)$, where $a_i \in \mathbb{C}$ are coordinates of the space.
Alternatively, a projective $n$-space $\mathbb{P}^n = \mathbb{C}^{n+1}$ is described by the $(n+1)$-tuples $(a_0, \dots , a_n)$ under the scaling relation $(a_0 , \dots , a_n) \sim ( \lambda a_0 , \dots , \lambda a_n)$, where $\lambda$ is a non-zero scalar and the point $(a_0 , \dots , a_n) = 0$ is excluded.\cite{Hartshorne}

An affine space can be viewed as the subset of a projective space where $a_0 \neq 0$.
In contrast, points where $a_0=0$ are referred to as ``points at infinity'' and allow geometric intersection results to be consistently defined without exceptions.
For example, in $\mathbb{A}^2$  two lines must always intersect exactly once unless they are parallel, whilst in $\mathbb{P}^2$ parallel lines intersect at a point at infinity.
Therefore, in the projective space  $\mathbb{P}^2$, the intersection rule is generalised without exceptions. 

\begin{figure*}[tbh!]
\input{figures/TikZ_circle_projection.tex}
\hspace{1em}\includegraphics[scale=1.25, trim={3.7cm 2cm 1.2cm 2.cm}, clip]{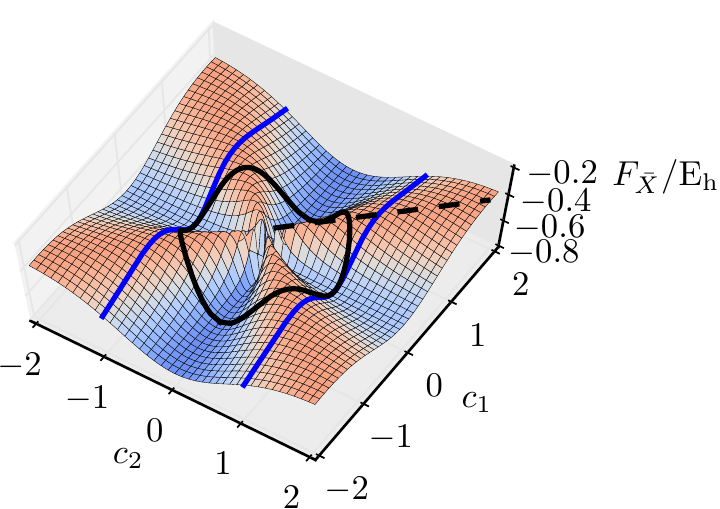}
\caption{
Constrained projective h-RHF energy $F_{\bar{X}} $ for the $n=2$ system \ce{H2} (STO-3G) at a bond length of $2.5$\r{A}.
The coordinate $c_0$ defines the normalisation constant, representing the distance of a point from the origin, and $\frac{\partial F_{\bar{X}}}{\partial c_0} = 0$ for all $c_0 \neq 0$.
Exploiting this invariance, every stationary point constrained to the circle $c_1^2 + c_2^2 = 1$ (black curve) with $c_2 \neq 0$ can also be located as a stationary point constrained to the line $c_2 = 1$ (blue line), where rescaling recovers the normalised h-RHF state (dashed line).
Due to the overall sign symmetry, only half the stationary points need to be considered (filled circles vs. open circles).  
}
\label{fig:n_2_solutions}
\end{figure*}

Using this terminology, the spatial orbital $\phi$ with $n$ basis functions is represented by a point $\left( c_1,\dots, c_n \right)$ in the affine space $\mathbb{A}^{n}=\mathbb{C}^{n}$.
The holomorphic Hartree--Fock energy (Equation \ref{eq:HoloEnergy}) is a function $\widetilde{E} : \mathbb{A}^{n} \rightarrow \mathbb{C}$ given by a polynomial of degree 4.
Satisfying the normalisation constraint (Equation \ref{eq:normalisation}) involves restricting solutions to the hypersurface $X \subseteq \mathbb{A}^n$ defined as
\begin{align}
\label{eq:AffineHypersurface}
X = \Big\lbrace (c_1, \dots, c_n) \in \mathbb{A}^{n} \mid \sum_{\mu=1}^n c_{\mu}^2 =1  \Big\rbrace .
\end{align}
Points corresponding to h-RHF states are then the vanishing points of the differential $\mathrm{d} \widetilde{E}$ restricted to $X$.

To enable a complete enumeration of these points, we must first convert to the projective space $\mathbb{P}^{n}$ represented by the points $\left( c_0,\dots,c_n \right)$.
This is achieved through the mapping $(c_1, \dots, c_n) \mapsto (\frac{c_1}{c_0}, \dots, \frac{c_n}{c_0})$, converting all polynomials in the affine coordinates $\{ c_1, \dots, c_n \}$ to homogeneous polynomials in the projective coordinates $\{ c_0, \dots, c_n \}$.
Following this transformation, the constraint becomes 
\begin{align}
\label{eq:ProjectiveConstraint}
\sum_{\mu=1}^n c_{\mu}^2 = c_0^2 
\end{align}
and solutions are therefore restricted to the hypersurface $\widebar{X} \subseteq \mathbb{P}^n$ defined as
\begin{align}
\label{eq:ProjectiveHypersurface}
\widebar{X} =  \Big\lbrace  (c_0,  \dots , c_n)  \in \mathbb{P}^{n} | \sum_{\mu=1}^n c_{\mu}^2 = c_0^2  \Big\rbrace .
\end{align}
The holomorphic energy can then be written as a rational function on $\mathbb{P}^n$
\begin{align}
\label{eq:ProjectiveHoloEnergy}
 F\left(c_0,\dots, c_n \right) &= \widetilde{E} \left( \frac{c_1}{c_0}, \dots, \frac{c_n}{c_0} \right) \nonumber \\
 &=  \frac{\widebar{E}\left(c_0,\dots, c_n \right)}{c_0^{4}},
\end{align}
where $\widebar{E}$ is the homogeneous version of $\widetilde{E}$ given by
\begin{align}
\label{eq:HomogeneousHoloEnergy}
\widebar{E}\left(c_0,\dots, c_n \right) &=  h_{0}^{\vphantom{4}} c_0^4 + 2 \sum_{\mu, \nu=1}^n h_{\mu \nu}^{\vphantom{4}} c_{\mu}^{\vphantom{4}} c_{\nu}^{\vphantom{4}} c_0^2  \nonumber \\
&+ \sum_{\mu, \nu, \sigma, \tau=1}^n h_{\mu \nu \sigma \tau} c_{\mu} c_{\nu} c_{\sigma} c_{\tau} .
\end{align}
Consequently, h-RHF states exist as vanishing points of the differential 
\begin{align}
\label{eq:ProjectiveDifferential}
\mathrm{d} F = \frac{\partial F}{\partial c_0} \mathrm{d} c_0 + \sum_{\mu = 1}^{n} \frac{\partial F}{\partial c_{\mu}} \mathrm{d} c_{\mu}
\end{align}
restricted to the hypersurface $\widebar{X}$.

From here, it can be shown that, including multiplicities, the number of such vanishing points (and thus the exact number of h-RHF solutions) is given by 
\begin{align}
\label{eq:NumberSolutions}
N_{\mathrm{solutions}} = \frac{1}{2} \left( 3^n - 1 \right).
\end{align}
A rigorous proof of this relationship is mathematically involved and beyond the scope of the current communication, but will form the focus of a future publication.
Instead, here we present a more intuitive derivation.

Consider the case of one basis function, $n=1$; clearly there are two trivial solutions at $(-1)$ and $(1)$ in the affine space $\mathbb{A}^1$.
Both points give the same density matrix and therefore describe equivalent h-RHF states.
This overall sign symmetry arises for all h-RHF states and is henceforth implicit.

\begin{figure*}[htb!]
\input{figures/OrbitalPlots.tex}
\caption{
Constrained projective h-RHF energy $F_{\bar{X}} $ plotted on the sphere $c_1^2 + c_2^2 + c_3^2 = 1$ and the plane $c_3=1$ for linear \ce{H3^+} (STO-3G) at a bond length of $2.5$\r{A}.
Nine stationary points can be located on the plane $c_3 = 1$, as shown alongside their corresponding orbital plots.
The remaining four h-RHF states exist at the infinities of this plane, and can be located by finding the stationary points constrained to the plane $c_3=0$.
The bond length has been chosen such that all solutions and their energies are real, although the results  extend to geometries where complex h-RHF states are present.
}
\label{fig:n_3_solutions}
\end{figure*}
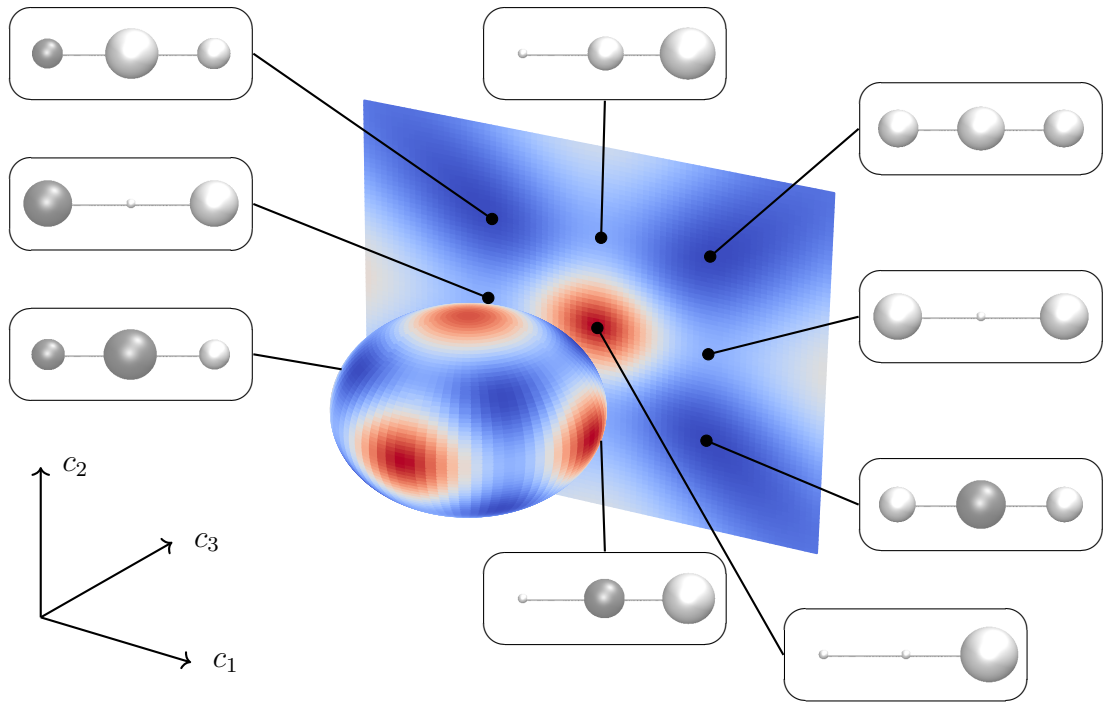 

Now consider the $n=2$ case, represented by a point $(c_0 , c_1 , c_2)$ in projective space $\mathbb{P}^2$. 
The projective h-RHF energy $F$ is then given by
\begin{align}
F \left( c_0, c_1, c_2 \right) &= h_{0} + 2 \sum_{\mu, \nu=1}^2 h_{\mu \nu} \frac{c_{\mu} c_{\nu}}{c_0^2}  \nonumber \\
&+ \sum_{\mu, \nu, \sigma, \tau =1}^2 h_{\mu \nu \sigma \tau} \frac{c_{\mu} c_{\nu} c_{\sigma} c_{\tau}}{c_0^4}.
\end{align}
Restriction to the hypersurface $\widebar{X}$, in this case given by $c_0^2 = c_1^2 + c_2^2$, causes $F$ to become equivalent to the normalised h-RHF energy where $c_0$ provides the normalisation factor.
Consequently, the constrained function $F_{\widebar{X}}$ is invariant to a global rescaling of the orbital coefficients $c_1$ and $c_2$ and the  partial derivative $\frac{\partial F_{\widebar{X}}}{\partial c_0}$ is zero for all $c_0 \neq 0$, as illustrated for \ce{H2} (STO-3G) in Figure \ref{fig:n_2_solutions}.
Although it is possible for solutions to exist with $c_0 = 0$, we find these arise only when electron-electron interactions vanish completely.
We do not expect this to occur in real molecular systems, and therefore continue our intuitive derivation under the assumption that $c_0 \neq 0$ for all stationary points.

First consider the case $c_2 \neq 0$.
Exploiting the invariance of $F_{\widebar{X}}$ to $c_0$ allows the h-RHF solutions to be located as stationary points along either the circle $c_1^2 + c_2^2 = 1$ (black curve in Figure \ref{fig:n_2_solutions}) or the line $c_2 = 1$ (blue line in Figure \ref{fig:n_2_solutions}).
Taking the latter approach enforces $\mathrm{d} c_2 = 0$ and, when combined with $ \frac{\partial F_{\widebar{X}}}{\partial c_0} = 0$, the constrained differential becomes 
\begin{align}
\label{eq:ProjetiveDifferential_n=2}
\mathrm{d} F_{\widebar{X}} =\left. \frac{\partial F_{\widebar{X}}}{\partial c_1}\right|_{c_2 = 1} \mathrm{d} c_1.
\end{align}
Since $F$ is a fourth degree polynomial in $c_1$, the partial derivative $\left. \frac{\partial F_{\widebar{X}}}{\partial c_1}\right|_{c_2 = 1}$ is third degree in $c_1$ and has three roots, each defining an h-RHF state.
Next we consider the case $c_2 = 0$, recovering the $n=1$ system and yielding one further solution in the affine space $\mathbb{A}^2$ at $(1,0)$.
The total number of solutions for two basis functions is therefore $ 3 + 1 = 4 $. 
 
We continue by adding a third basis function, represented in $\mathbb{P}^3$ by the point $(c_0 , c_1 , c_2 , c_3)$, and rotate the orbital coefficients such that $\frac{\partial F_{\widebar{X}}}{\partial c_3} = 0$ wherever $c_3=0$.
First consider $c_3 \neq 0$.
Similarly to $n=2$, the h-RHF states can be located as stationary points on either the sphere $c_1^2 + c_2^2 + c_3^2 = 1$ or the plane $c_3 = 1$, as shown for \ce{H3^+} (STO-3G) in Figure \ref{fig:n_3_solutions}.
By considering the stationary points on the plane $c_3 = 1$, the constrained differential $\mathrm{d} F_{\widebar{X}}$  reduces to
\begin{align}
\label{eq:ProjetiveDifferential_n=3}
\mathrm{d} F_{\widebar{X}} = \left. \frac{\partial F_{\widebar{X}}}{\partial c_1}\right|_{c_3 = 1} \mathrm{d} c_1 + \left. \frac{\partial F_{\widebar{X}}}{\partial c_2}\right|_{c_3 = 1} \mathrm{d} c_2.
\end{align}
The required solutions are now located by finding the common intersections of the third degree homogeneous polynomials
\begin{align}
\left. \frac{\partial F_{\widebar{X}}}{\partial c_1}\right|_{c_3 = 1} = 0 \hspace{1em} \mathrm{and} \hspace{1em} 
\left. \frac{\partial F_{\widebar{X}}}{\partial c_2}\right|_{c_3 = 1} = 0.
\label{eq:Intersections_n=3}
\end{align} 
B\'{e}zout's Theorem states that the number of common intersections of $n$ homogeneous polynomials in the projective space $\mathbb{P}^{n}$ is given by the product of the degrees of each polynomial.\cite{Hartshorne}
Consequently, the number of solutions to (\ref{eq:Intersections_n=3}) is given by $3 \times 3 = 9$, yielding nine h-RHF states with $c_3 \neq 0$.
We continue by considering the case where $c_3 = 0$ and recover a system of two basis functions analogous to Figure \ref{fig:n_2_solutions}.
This regime yields a further $3 + 1 = 4$ solutions, and thus the total number of h-RHF states for $n=3$ is $9 + 3 + 1 = 13$.

We can iteratively extend this argument to a general two--electron system with $n$ basis functions and find the number of solutions is given by
\begin{align}
\label{eq:NumberSolutionsIntuitive}
N_{\mathrm{solutions}} &= \sum_{i=0}^{n-1} 3^{i}.
\end{align} 
Expressing this geometric series in a closed form then recovers Equation \ref{eq:NumberSolutions}.
Crucially, both this intuitive derivation and the more rigorous proof are independent of the nuclear repulsion, one- and two-electron integrals.
Therefore, the number of h-RHF solutions  depends only on the number of basis functions and every solution must be conserved as the geometry or atomic charges of a system are varied.

It is important to note that Equation \ref{eq:NumberSolutions} may include solutions with a multiplicity greater than one, for example exactly when states coalesce at the Coulson--Fischer point.
Alternatively, it is possible for continuous lines or planes of solutions in the orbital coefficient space to exist.
We believe this will occur for systems with degenerate basis functions, for example molecules with cylindrical symmetry, however an infinite number of solutions can be avoided by forcing the single-particle orbitals to transform as an irreducible representation of the molecular point group.

We also note that Equation \ref{eq:NumberSolutions} has previously been identified by Stanton as an upper bound on the number of real closed-shell Hartree--Fock solutions for two-electron systems.\cite{Stanton1968}
Stanton arrived at this result geometrically for the $n=2$ case, but was restricted to considering the $n\geq 3$ case in the zero differential overlap limit, where
\begin{align}
\label{eq:ZeroDiffOvA}
h_{\mu \nu} = \delta_{\mu \nu} h_{\mu \mu}
\end{align}
and
\begin{align}
\label{eq:ZeroDiffOvB}
(\mu \nu | \sigma \tau ) = \delta_{\mu \nu} \delta_{\sigma \tau} (\mu \mu | \sigma \sigma ).
\end{align}
In contrast, employing the algebraic geometry approach presented above yields an entirely generalised geometric proof.
Furthermore, our approach proves that Equation \ref{eq:NumberSolutions} provides not only an upper bound on the number of real RHF solutions, but also the \textit{exact} number of h-RHF states for two--electron systems.
We believe that using algebraic geometry will subsequently enable the number of holomorphic solutions to be computed for both unrestricted or many electron systems, however there may be challenges in obtaining a general closed formula for these cases.

Finally, we note that the number of h-RHF states predicted by Equation \ref{eq:NumberSolutions} is much larger than the dimension of the full configuration interaction (FCI) space for two-electrons, given by $n^2$.
However, the non-orthogonality of different SCF solutions allows each state to span multiple FCI determinants, enabling a more compact description of the Hilbert space through a small number of relevant h-RHF states.

\section{Closed-shell states of \ce{HZ} in STO-3G}
\label{Sect:HZ}

\begin{figure*}[!htb]
    \includegraphics[scale=1.0, trim={0.7cm 0.1cm 0.8cm 0.2cm}, clip]{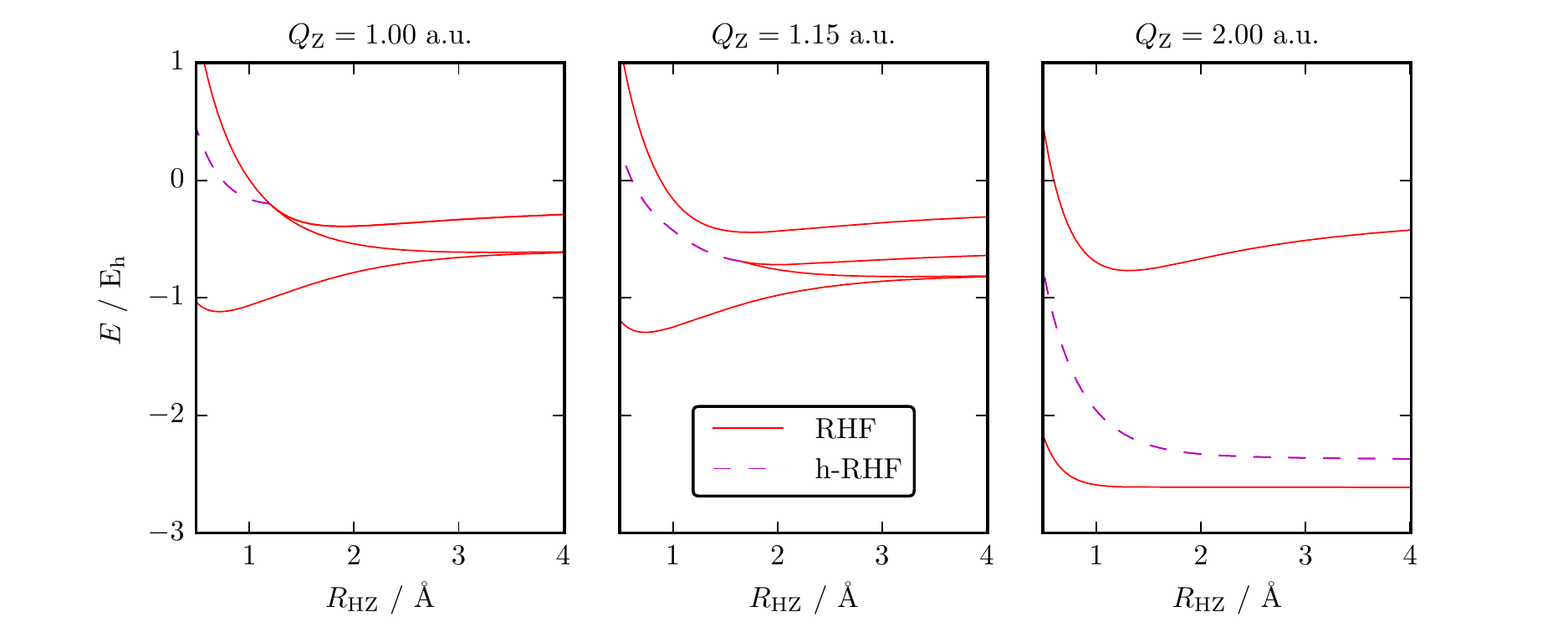}
\caption{Four h-RHF states for \ce{HZ} (STO--3G) are located for all bond lengths and charges $Q_{\mathrm{Z}}$.
Each h-RHF state corresponds to a real RHF state (red/solid) where such states exist.
h-RHF states with complex orbital coefficients (magenta/dashed) form in complex--conjugate pairs with degenerate standard Hartree--Fock energies.
In the case of $Q_{\mathrm{Z}}=2.00 \ \mathrm{a.u.}$, corresponding to \ce{HHe^+} a pair of degenerate states exist with complex coefficients across all geometries.}
\label{fig:HZ_plotVsR}
\end{figure*}

Since the number of h-RHF states for two-electron systems depends on only the number of basis functions, any pair of distinct two-electron systems with the same number of basis functions can be smoothly interconverted by either moving the basis function centres (eg. changing structure) or adjusting the atomic charges (eg. changing atoms).
This concept can be demonstrated by considering the h-RHF solutions of the \ce{HZ} molecule using the STO-3G basis set.
Varying the nuclear charge of the hydrogenic \ce{Z} atom, $Q_{\mathrm{Z}}$, between $0$ and $2$, enables the smooth interconversion along the isoelectronic sequence \ce{H^-} $\rightarrow$ \ce{H2} $\rightarrow$ \ce{HHe^+}.\cite{King1969}
This simple system is of particular interest as an archetypal model for the qualitative nature of the h-RHF states in symmetric and asymmetric diatomics.

The sole occupied spatial orbital is expressed in terms of the RHF rotation angle $\theta$ describing the degree of mixing between the $1\mathrm{s}$ atomic orbitals on H and Z,
\begin{align}
\phi \left( \mathbf{r} \right) &= \sin \left( \theta - \frac{\pi}{4} \right) \chi_{\mathrm{1s,H}} \left( \mathbf{r} \right) \nonumber \\
&+ \cos \left( \theta - \frac{\pi}{4} \right) \chi_{\mathrm{1s,Z}} \left( \mathbf{r} \right).
\label{eq:RotatingBasisFunctions}
\end{align}
With two basis functions, Equation \ref{eq:NumberSolutions} dictates that four h-RHF states exist for all bond lengths $R_{\mathrm{HZ}}$ and values of $Q_{\mathrm{Z}}$.

\begin{figure*}[!bt]
  \centering
\begin{subfigure}{.49\textwidth}
  \centering
  \includegraphics[scale=1, trim={0.7cm 0.cm 1.5cm 1.1cm}, clip]{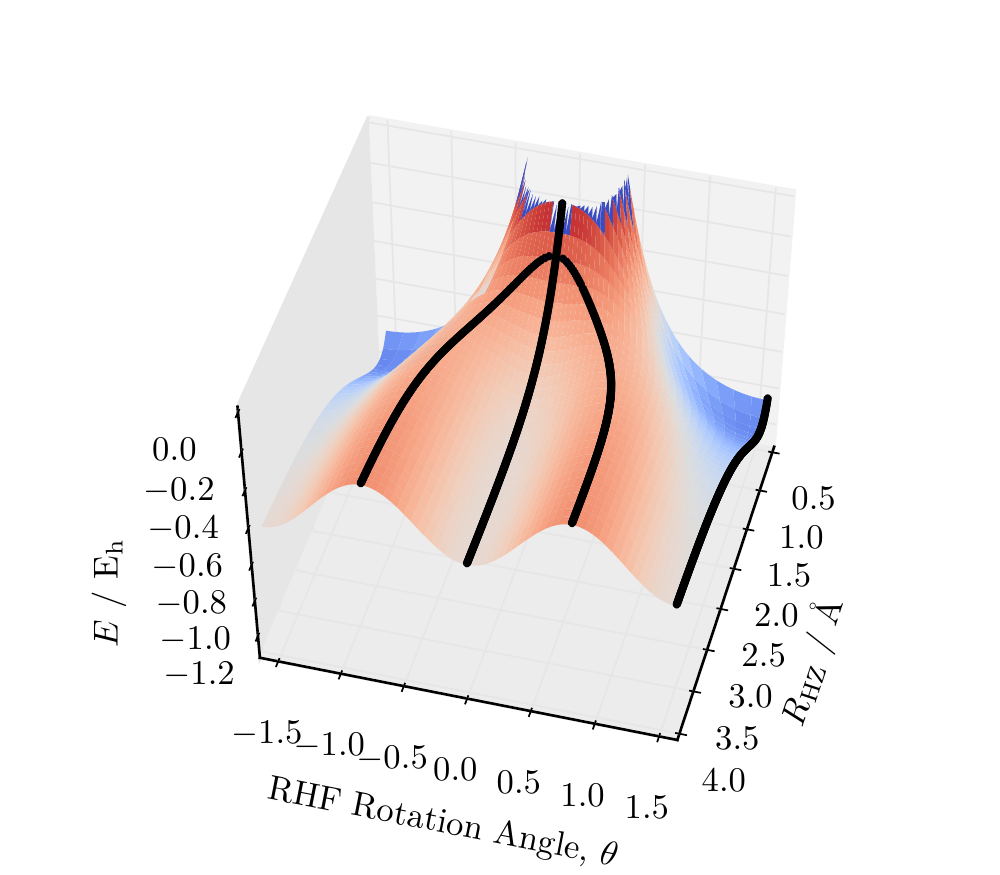}
  \caption{}
  \label{fig:HZ_surface_z1-00}
\end{subfigure}
\begin{subfigure}{.49\textwidth}
  \centering
  \includegraphics[scale=1., trim={0.7cm 0.cm 1.5cm 1.1cm}, clip]{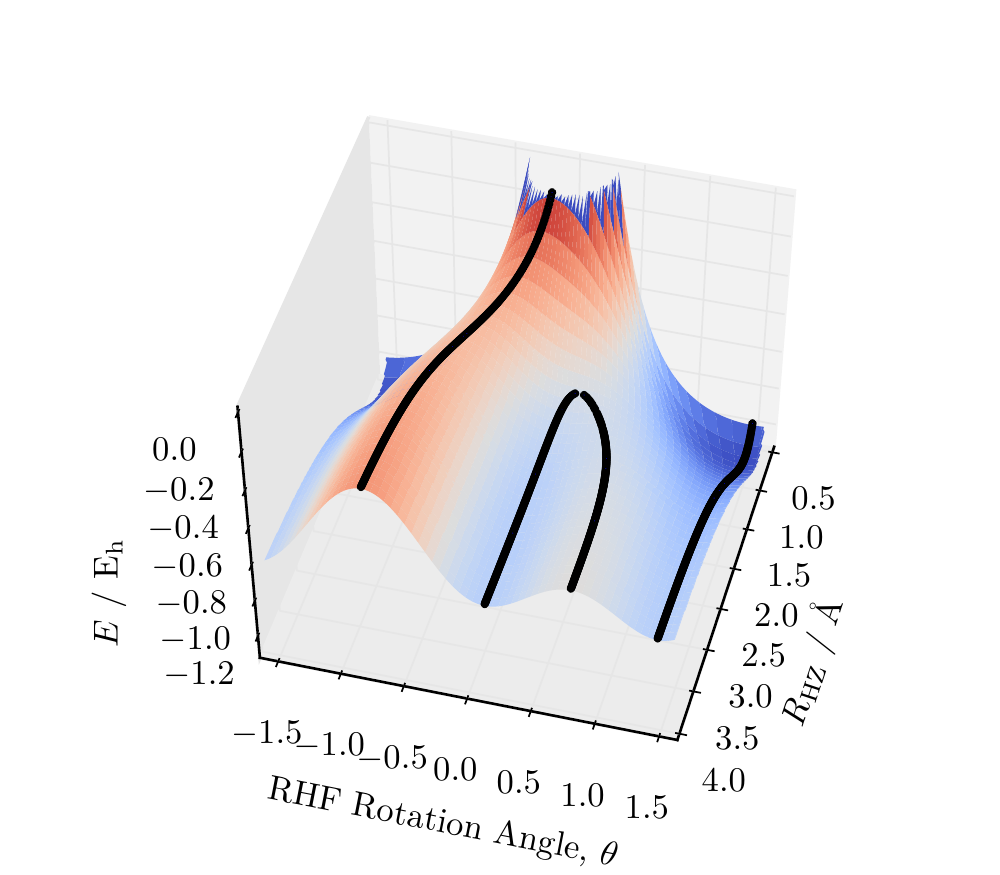}
  \caption{}  
  \label{fig:HZ_surface_z1-15}
\end{subfigure}
\caption{Conventional Hartree--Fock energy plotted as a function of the RHF rotation angle, $\theta$ (Equation \ref{eq:RotatingBasisFunctions}) for $Q_{\mathrm{Z}} = 1.00\ \mathrm{a.u.}$ and $Q_{\mathrm{Z}} = 1.15\ \mathrm{a.u.}$.
\subref{fig:HZ_surface_z1-00} When $Q_{\mathrm{Z}} = 1.00\ \mathrm{a.u.}$ the molecule possesses $\mathcal{D}_{\infty h}$ symmetry and the ionic solutions simultaneously converge with the $\upsigma_{\mathrm{u}}^2$ state, disappearing at a triply degenerate $\mathrm{A_{3}}$ cusp catastrophe in a pitchfork bifurcation.
\subref{fig:HZ_surface_z1-15} For $Q_{\mathrm{Z}} = 1.15\ \mathrm{a.u.}$ the molecular symmetry becomes $\mathcal{C}_{\infty v}$, decomposing the pitchfork bifurcation into a primary branch and two secondary modes that coalesce and disappear at a doubly degenerate $\mathrm{A_{2}}$ fold catastrophe.}
\label{fig:HZ_surfaces}
\end{figure*}

We begin by considering the case where $Q_{\mathrm{Z}} = 1.00 \ \mathrm{a.u.}$, corresponding to \ce{H2}, and plot the conventional Hartree--Fock energy of each h-RHF solution from $R_{\mathrm{HZ}}=0.5$ \r{A} to $R_{\mathrm{HZ}}=4.0$ \r{A} in Figure \ref{fig:HZ_plotVsR} (left panel).
In the dissociation limit, each solution has real orbital coefficients and corresponds to a real Hartree--Fock state (red/solid), representing the $\upsigma_\mathrm{g}^2$, $\upsigma_{\mathrm{u}}^2$ and degenerate ionic \ce{H+-Z-} and \ce{H^{-}-Z+} states in order of ascending energy.
As the internuclear distance is reduced, the ionic states coalesce with the $\upsigma_{\mathrm{u}}^2$ state at a Hartree--Fock instability threshold and disappear at shorter bond lengths.
In contrast, the corresponding h-RHF solutions continue to exist with complex orbital coefficients (magenta/dashed), forming a degenerate pair related by complex conjugation.
Significantly, although the conventional Hartree--Fock energy of these states appears kinked at the coalescence point, their path through orbital coefficient space is both smooth and continuous, and it is this property that is essential for NOCI. 
Using the classification of Hartree--Fock singlet instability thresholds developed by Mestechkin\cite{Mestechkin1978,Mestechkin1979,Mestechkin1988}, the coalescence point for $Q_{\mathrm{z}} = 1.00\ \mathrm{a.u.}$ can be identified as a ``confluence'' point, where two maxima converge onto a minimum, as shown in Figure \ref{fig:HZ_surface_z1-00}.

In contrast, the molecular symmetry is broken by moving to $Q_{\mathrm{Z}} = 1.15\ \mathrm{a.u.}$ (middle panel of Figure \ref{fig:HZ_plotVsR}), lifting the degeneracy of the ionic states and leading to the coalescence of only the $\upsigma_{\mathrm{u}}^2$ and \ce{H+ - Z^-} states at a ``pair annihilation'' point.
Beyond this point, both real RHF solutions disappear whilst, again, their h-RHF counterparts continue as a complex degenerate pair.
The existence of complex h-RHF states arising at this pair annhilation point indicates the applicability of holomorphic Hartree--Fock for vanishing states in asymmetric diatomics including \ce{LiF}\cite{Thom2009}.

\begin{figure}[!hb]
\center 
\includegraphics[scale=0.95, trim={0.5cm 0.2cm 0.8cm 0.5cm},clip]{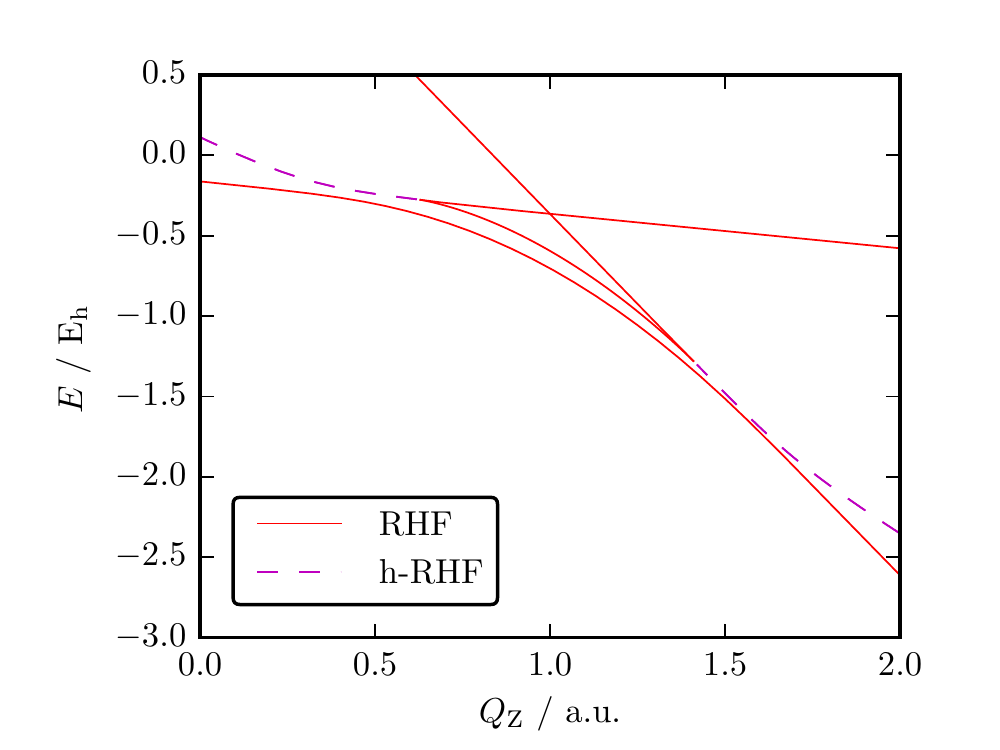}
\caption{
Four h-RHF solutions of \ce{HZ} (STO--3G) are plotted against the nuclear charge $Q_{\mathrm{Z}}$ for a bond length of $2.50$\r{A}, showing the smooth interconversion between the complex h-RHF states of \ce{HHe^+} (at $Q_{\mathrm{Z}}=2.00\ \mathrm{a.u.}$) and \ce{H2} (at $Q_{\mathrm{Z}}=1.00\ \mathrm{a.u.}$).
Two h-RHF states of \ce{HHe^+} that exist with complex coefficients for all bond lengths are seen to smoothly interconvert with real h-RHF states of \ce{H2} as $Q_{\mathrm{Z}}$ is varied.
}
\label{fig:HZ_charge_250}
\end{figure}

As $Q_{\mathrm{Z}}$ is increased further, the coalescence point occurs at increasing bond lengths until eventually only two real Hartree--Fock solutions exist across all geometries, as demonstrated for $Q_{\mathrm{Z}} = 2.00\ \mathrm{a.u.}$ (right panel of Figure \ref{fig:HZ_plotVsR}).
The remaining two h-RHF solutions form a degenerate pair with complex orbital coefficients across all geometries.
Although no electronic state appears to correspond to these complex solutions, they can be smoothly evolved into real states with physical significance by varying $Q_{\mathrm{Z}}$, as shown in Figure \ref{fig:HZ_charge_250}.
Consequently, we believe these states should be considered as `dormant' analytic continuations of real states.

The field of catastrophe theory allows the nature of real RHF coalescence points in \ce{HZ} to be comprehensively understood as both $Q_{\mathrm{Z}}$ and $R_{\mathrm{HZ}}$ are varied.
Catastrophe theory provides a framework for qualitatively investigating stationary points for potentials that depend on a certain set of system control parameters.\cite{Gilmore}
Generally, applications focus on degenerate equilibrium points where one or more higher derivatives of the potential function are zero, referred to as non-Morse critical points.
Expanding the potential at these points as a Taylor series in small perturbations of the parameters allows the degeneracy to be lifted in a process referred to as ``unfolding''.
For one-dimensional potentials, this allows any non-Morse critical point to be classified as one of only seven ``elementary catastrophes''\cite{Thom}.

In \ce{HZ}, the number of stationary points of the conventional Hartree--Fock energy is controlled by two physical parameters $R_{\mathrm{HZ}}$ and $Q_{\mathrm{Z}}$, and we consider the behaviour of stationary points around $R_{\mathrm{HZ}} = 1.19$ \r{A}, $Q_{\mathrm{Z}} = 1.00\ \mathrm{a.u.}$ and $\theta = 0$, corresponding to the RHF confluence point of \ce{H2}.
When $Q_{\mathrm{Z}} = 1.00\ \mathrm{a.u.}$, the confluence point is a triply degenerate non-Morse critical point and the RHF solutions disappear in a pitchfork bifurcation as shown in Figure \ref{fig:HZ_surface_z1-00}. 
In contrast, for $Q_{\mathrm{Z}} \neq 1.00\ \mathrm{a.u.}$ the pair annihilation point corresponds to a doubly degenerate non-Morse critical point and the pitchfork bifurcation is broken into a primary branch, existing across all geometries, and two secondary solutions which coalesce and disappear, as shown in Figure \ref{fig:HZ_surface_z1-15}. 

Simultaneously considering the stationary points as both  $R_{\mathrm{HZ}}$ and $Q_{\mathrm{Z}}$ are varied reveals the related elementary catastrophe to be a triply degenerate cusp or $\mathrm{A}_3$ catastrophe.\cite{Gilmore}
In contrast, pair annihilation points correspond to doubly degenerate fold or $\mathrm{A}_2$ catastrophe.
This identification indicates fold catastrophes are significantly more widespread than cusp catastrophes in molecular systems, with the simultaneous convergence of three RHF states in \ce{H2} arising directly from the additional plane of symmetry.
Despite this, the existence of complex h-RHF solutions for each case in Figure \ref{fig:HZ_plotVsR} demonstrates that holomorphic Hartree--Fock states will always exist regardless of the molecular symmetry or the nature of the singlet instability.

\section{Isomorphism of \ce{HHeH^2+} and \ce{HHeH}}
\label{Sect:Isomorphism}

\begin{figure*}[htb!]
 \centering 
 \begin{subfigure}{.49\textwidth}
  \flushleft
  \includegraphics[scale=1.0, trim={0.0cm 0.0cm 0.6cm 0.5cm},clip]{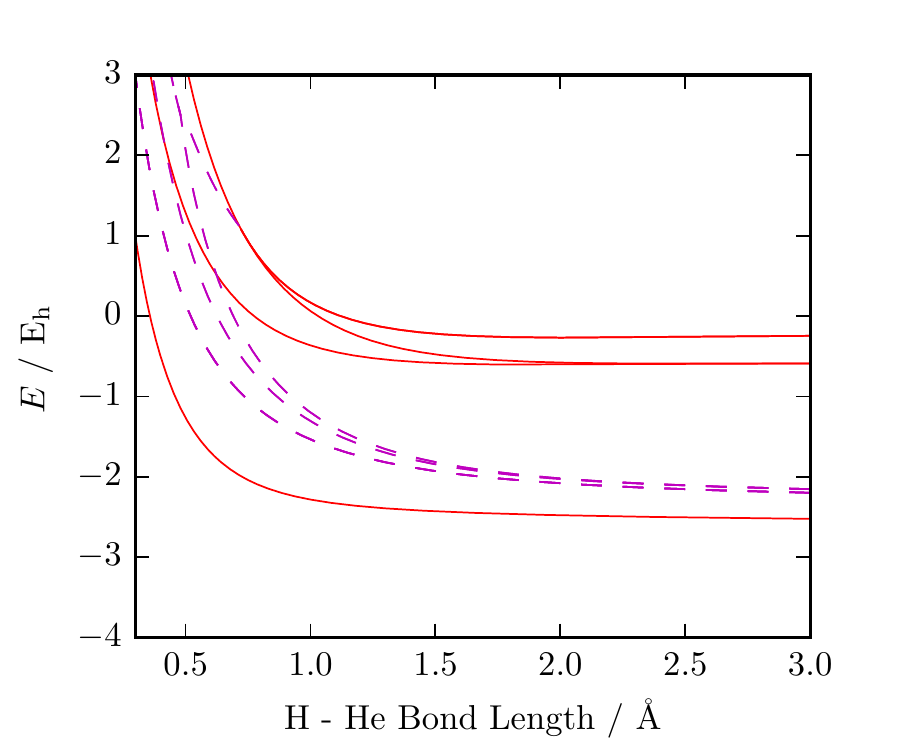}
  \caption{\ce{HHeH^{2+}}}
  \label{fig:H2He_2p}
 \end{subfigure}
 \hspace{-1.4cm}
 \begin{subfigure}{.49\textwidth}
  \flushright
  \includegraphics[scale=1.0, trim={1.0cm 0.0cm 0.6cm 0.5cm},clip]{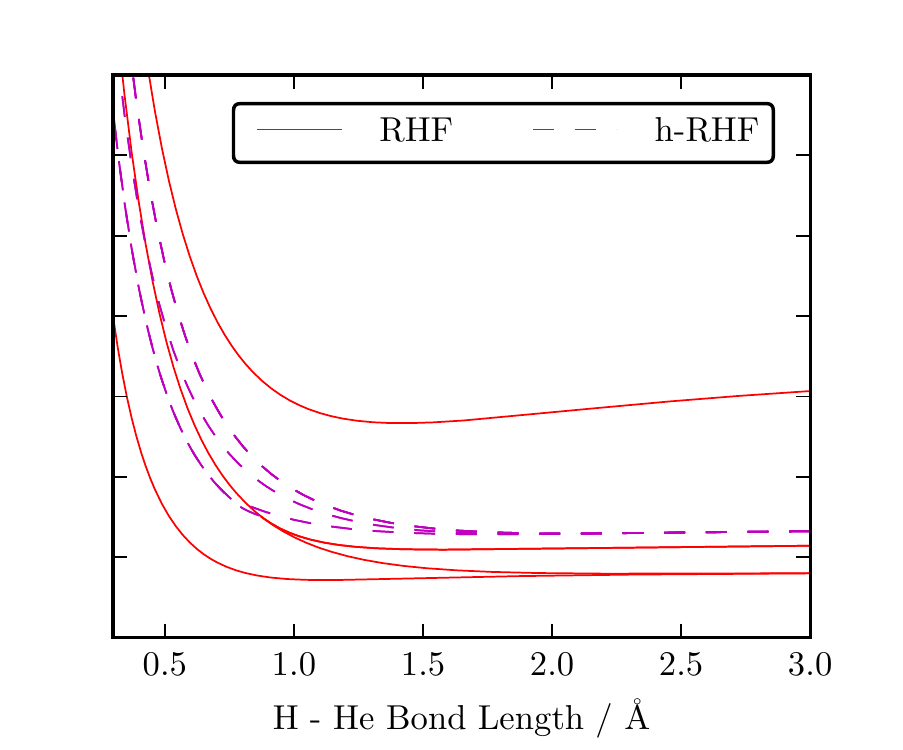}
  \caption{\ce{HHeH}}
  \label{fig:H2He}
 \end{subfigure}
 \caption{The 13 h-RHF states for \subref{fig:H2He_2p} \ce{HHeH^{2+}} and \subref{fig:H2He} \ce{HHeH} in STO-3G for a range of symmetric \ce{H-He} bond lengths.
 Six solutions have complex coefficients across all bond lengths, arising in degenerate complex-conjugate pairs.
 At around $0.7$ \r{A} in \ce{HHeH^2+} and around $0.9$ \r{A} in \ce{HHeH}, two pairs of complex solutions coalesce to form a four-fold degenerate set of complex h-RHF states.}
 \label{fig:H2He_comparison}
\end{figure*}

We now consider the two--electron linear \ce{HHeH^2+} molecule using the STO-3G basis set.
As a system with three basis functions, Equation \ref{eq:NumberSolutions} predicts 13 h-RHF states, plotted across a range of symmetric bond lengths in Figure \ref{fig:H2He_2p}.
Similarly to \ce{H2}, \ce{HHeH^2+} possesses $\mathcal{D}_{\infty \mathrm{h}}$ symmetry and thus the disappearance of the high energy real RHF states occurs at a triply degenerate cusp $\mathrm{A}_3$ catastrophe.
Beyond this point, the corresponding h-RHF states become complex, forming a degenerate pair related by complex conjugation.
Eight further dormant solutions similar to those seen in \ce{HHe^+} exist with complex coefficients across all geometries.
Furthermore, at $R = 0.5$ \r{A} we observe the convergence of two pairs of degenerate complex h-RHF states to form a set of four degenerate complex solutions.

Mathematically, systems with two electrons or two electron holes in $n$ basis functions are isomorphic and have the same number of h-RHF states.
The \ce{HHeH^2+} and \ce{HHeH} systems in STO-3G provide one of the simplest example, as shown in Figure \ref{fig:H2He_comparison}.
In both cases there are 13 h-RHF states across the all molecular geometries including eight dormant states.
Again, in \ce{HHeH} the coalescence of two pairs of degenerate complex h-RHF states to form a set of four degenerate complex solutions can be observed at $R~=~0.7$~\r{A}.
Although the relative standard Hartree--Fock energies of the states in \ce{HHeH} are in the reverse order to those in \ce{HHeH^2+}, the qualitative behaviour of solutions at coalescence points is equivalent and arises between the same pairs of h-RHF states.

Exploiting this isomorphism allows Equation \ref{eq:NumberSolutions} to be extended to systems with $2n~-~2$ electrons.
To our knowledge, only Fukutome has previously attempted to enumerate the Hartree--Fock states for a general multiple electron system.\cite{Fukutome1971}
Fukutome expressed the Hartree--Fock problem as a density matrix equation to obtain lower and upper bounds on the number of complex Hartree--Fock solutions as $2^K$ and $2^{\left( n^2- K \right)}$, where $K = \min \left(N, n - N \right)$.
To represent closed-shell systems with two-electron holes we take $N=n-1$ and $K=1$, and thus Fukutome's result predicts lower and upper bounds of $2$ and $2^{(n^2-1)} = \frac{1}{2} \times 4^n$ respectively.
Since all real Hartree--Fock solutions are simultaneously also complex and holomorphic Hartree--Fock solutions, both Fukutome's expression and Equation \ref{eq:NumberSolutions} provide independent upper bounds on the number of real RHF states.
Consequently, our result of $\frac{1}{2} \times (3^n - 1)$ provides a significantly reduced upper bound on the number of real RHF states in these systems.

\section{Rotation of ethene}

\begin{figure*}[htb!]
 \centering 
 \begin{subfigure}{.49\textwidth}
  \flushleft
  \includegraphics[scale=1.0, trim={0.2cm 0.0cm 0.0cm 0.0cm},clip]{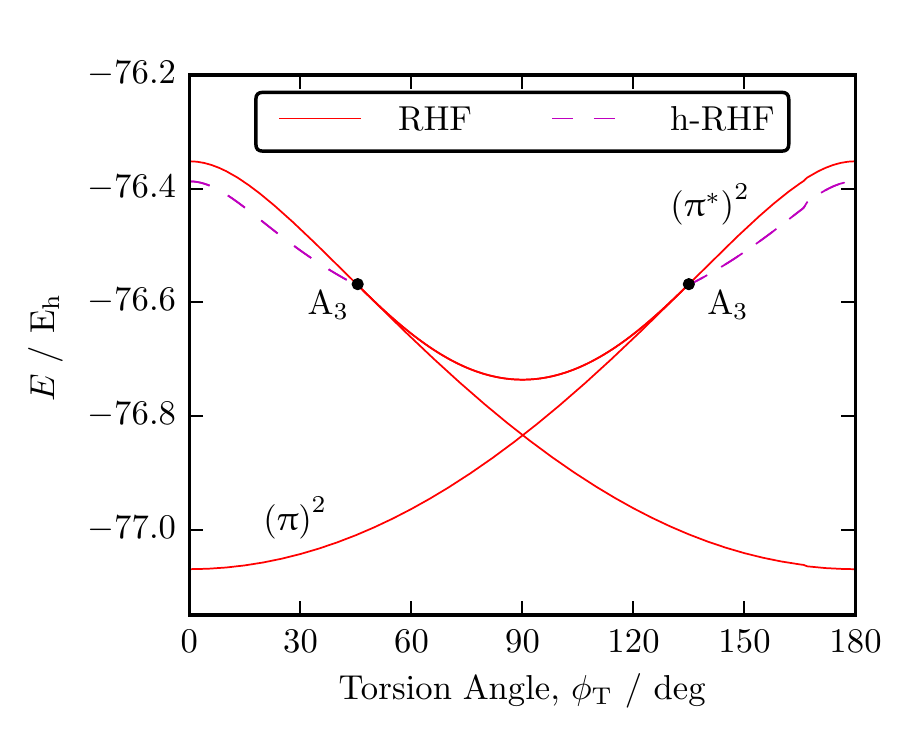}
  \caption{$Q_{\mathrm{X}}=6.0$\ a.u. (symmetric ethene)}
  \label{fig:C2H4}
 \end{subfigure}
 \begin{subfigure}{.49\textwidth}
  \flushright
  \includegraphics[scale=1.0, trim={1.4cm 0.0cm 0.0cm 0.0cm},clip]{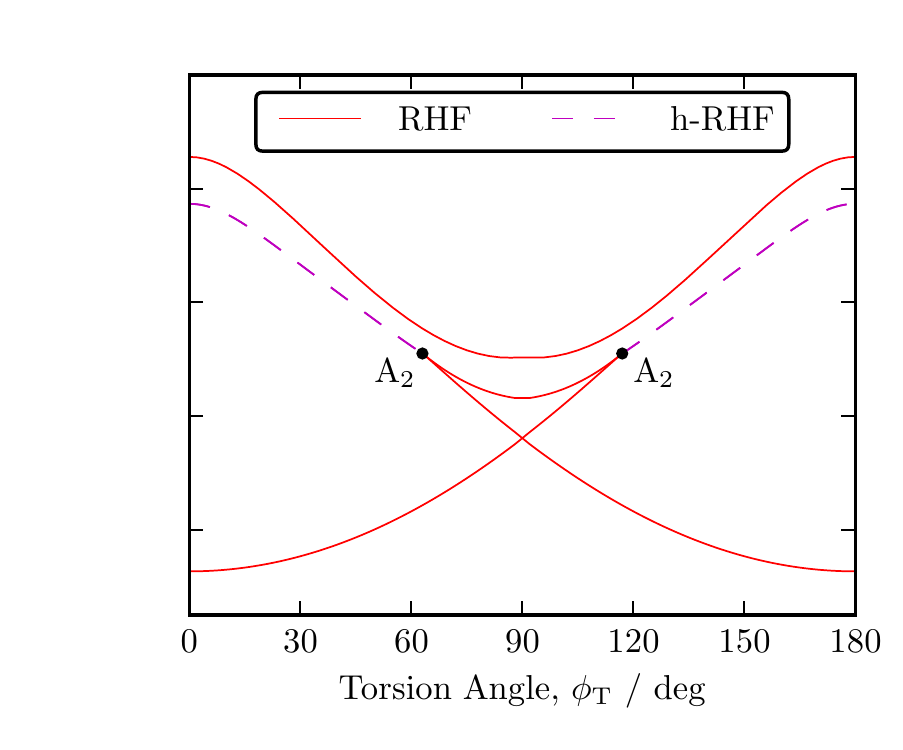}
  \caption{$Q_{\mathrm{X}}=6.1$\ a.u. (asymmetric ethene)}
  \label{fig:CZH4}
 \end{subfigure}
 \caption{The four h-RHF states for two electrons in the space of $\uppi$ and $\uppi^*$ orbitals of \ce{CH2XH2} (STO-3G) freezing the remaining core electrons and vitual orbitals at a \ce{C-X} bond length of $R_{\mathrm{CC}} = 1.256$ \AA. 
 \ce{X} is a carbon-like atom with nuclear charge $Q_{\mathrm{X}}$. 
 At $\phi_{\mathrm{T}} = 90^{\circ}$, every h-RHF solution corresponds to a real RHF state (red/solid).
 As $\phi_{\mathrm{T}}$ moves towards $0^{\circ}$ or $180^{\circ}$, real RHF states coalesce whilst h-RHF solutions continue with complex orbital coefficients (magenta/dashed).
 Breaking the molecular symmetry splits the degeneracy of the high energy ionic states, changing the coalescence points from triply degenerate $\mathrm{A_{3}}$ cusp catastrophes to doubly degenerate $\mathrm{A_{2}}$ fold catastrophes.}
 \label{fig:C2H4_r125_aTwist}
\end{figure*}

Although the examples presented in Sections \ref{Sect:HZ} and \ref{Sect:Isomorphism} provide insightful models for understanding the emergence of h-RHF solutions, we are not restricted to molecular systems containing only two electrons.
The properties and reactivity of many molecules are dominated by a subset of only two electrons which, by freezing the remaining core electrons, can also be considered as two-electron problems.
The electronic energy levels in the rotation of ethene, for example, depend strongly on the two-electron, two-centre $\uppi$ bond.

Starting with an orthogonal basis set composed of the STO-3G  ground state RHF molecular orbitals at the optimised planar $\mathcal{D}_{\mathrm{2h}}$ geometry, we select the $\mathrm{b_{3u}}$ ($\uppi$) and $\mathrm{b_{2g}}$ ($\uppi^*$) orbitals as an active pair and freeze the remaining core electrons and virtual orbitals.
An h-RHF calculation using the $\uppi$ electrons in this active space reduces the system to a two-electron problem in two basis functions, yielding 4 solutions through Equation \ref{eq:NumberSolutions}. 
Due to the symmetry equivalence of the carbon centres, the h-RHF states resemble those of \ce{H2}, corresponding at dissociation to the $\left( \uppi \right)^2$ and $\left( \uppi^* \right)^2$ configurations and the degenerate symmetry broken \ce{H2C^{+}-C^{-}H2} and \ce{H2C^{-}-C^{+}H2}  ionic states.
As the carbon-carbon  bond length $R_{\mathrm{CC}}$ is shortened, the ionic states coalesce with the $\left( \uppi^* \right)^2$ state at around $R_\mathrm{CC} = 1.29$ \AA\ in a triply degenerate $\mathrm{A_3}$ cusp catastrophe analogous to Figure \ref{fig:HZ_surface_z1-00}.

\begin{figure*}[!tb]
\input{figures/LipsBifurcation.tex} 
\caption{Sketch of the critical manifold (left) showing the types of coalescence points between real RHF solutions in ethene and their dependence on the molecular control parameters $R_{\mathrm{CC}}$, $\phi_{\mathrm{T}}$ and $Q_{\mathrm{X}}$.
Sections through the critical manifold (right) demonstrate the dependence of these coalescence points on $\phi_{\mathrm{T}}$ and $Q_{\mathrm{X}}$ at various values of  $R_{\mathrm{CC}}$.
Doubly and triply degenerate coalescence points correspond to $\mathrm{A_2}$ fold and $\mathrm{A_3}$ cusp catastrophes respectively. 
Two cusp catastrophes emerge from an $\mathrm{A_3^+}$ cusp creation catastrophe and recombine at an $\mathrm{A_3^-}$ cusp annihilation catastrophe.
Within the conoidal structure (shaded) there exist four real RHF states whilst outside there are only two.}
\label{fig:C2H4_critical_manifold}
\end{figure*}
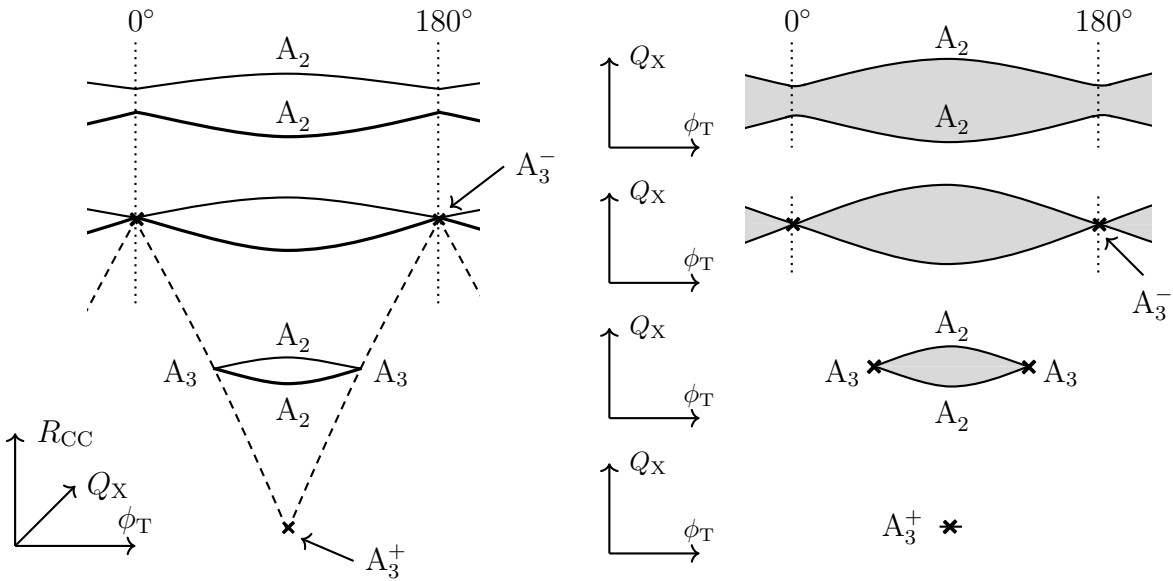

The evolution of the h-RHF states as the torsion angle $\phi_{\mathrm{T}}$ varies for $R_{\mathrm{CC}}=1.256$~\r{A} is shown in Figure \ref{fig:C2H4}. 
As $\phi_{\mathrm{T}}$ increases or decreases from the $90^{\circ}$ perpendicular structure ($\mathcal{D}_{\mathrm{2d}}$) towards the planar geometry, the ionic states simultaneously coalesce with the anti-bonding $\left( \uppi^* \right)^2$ state at two $\mathrm{A_3}$ cusp catastrophes located at around $\phi_{\mathrm{T}}=45^{\circ}$ and $135^{\circ}$.
This mirrors a previous analysis by Fukutome.\cite{Fukutome1973}
Beyond these singlet instability points, the related h-RHF states continue to exist with complex orbital coefficients.

Similarly to \ce{HZ}, breaking of the molecular symmetry can be modelled by replacing one carbon with a carbon-like nucleus \ce{X} containing six electrons and a variable nuclear charge $Q_{\mathrm{X}}$.
Increasing $Q_{\mathrm{X}}$ from $6.0$ a.u. splits the degeneracy of the ionic states, leading to the coalescence of only the $\left( \uppi^* \right)^2$ and \ce{H2X^{-}-C^{+}H2} states at two doubly degenerate $\mathrm{A_2}$ fold catastrophes that shift towards $\phi_{\mathrm{T}}=90^{\circ}$, as shown for $Q_{\mathrm{X}}=6.1$~a.u. in Figure \ref{fig:CZH4}.

The critical manifold, sketched in Figure \ref{fig:C2H4_critical_manifold}, demonstrates the evolution of these coalescence points over all possible variations of $R_{\mathrm{CC}}$, $\phi_{\mathrm{T}}$ and $Q_{\mathrm{X}}$.
For very short $R_{\mathrm{CC}}$ there exist only two real RHF states for all $\phi_{\mathrm{T}}$ and $Q_{\mathrm{X}}$.
As the bond length increases, two $\mathrm{A_3}$ cusp catstrophes emerge in the plane $Q_{\mathrm{X}}=6$ a.u.  from an $\mathrm{A_3^+}$ cusp creation point.\cite{Hidding2014}
These symmetry related  $\mathrm{A}_3$ catastrophes are connected by two $\mathrm{A}_2$ fold catastrophes when $Q_{\mathrm{X}} \neq 6$ a.u.
Further increasing $R_{\mathrm{CC}}$ causes the $\mathrm{A_3}$ catastrophes to move away from $\phi_{\mathrm{T}}=90^{\circ}$ until they recombine at $\phi_{\mathrm{T}}=0^{\circ}$ (or the symmetry related point $\phi_{\mathrm{T}}=180^{\circ}$) at an $\mathrm{A_{3}^{-}}$ cusp annhilation point\cite{Hidding2014}.
At larger bond lengths there are no coalescence points for $Q_{\mathrm{X}}=6$ a.u. whilst the two $\mathrm{A}_2$ fold catastrophes remain when $Q_{\mathrm{X}} \neq 6$ a.u.

\begin{figure*}[!htb]
\includegraphics[scale=1.0, trim={0.2cm 0.5cm 0.0cm 1.0cm},clip]{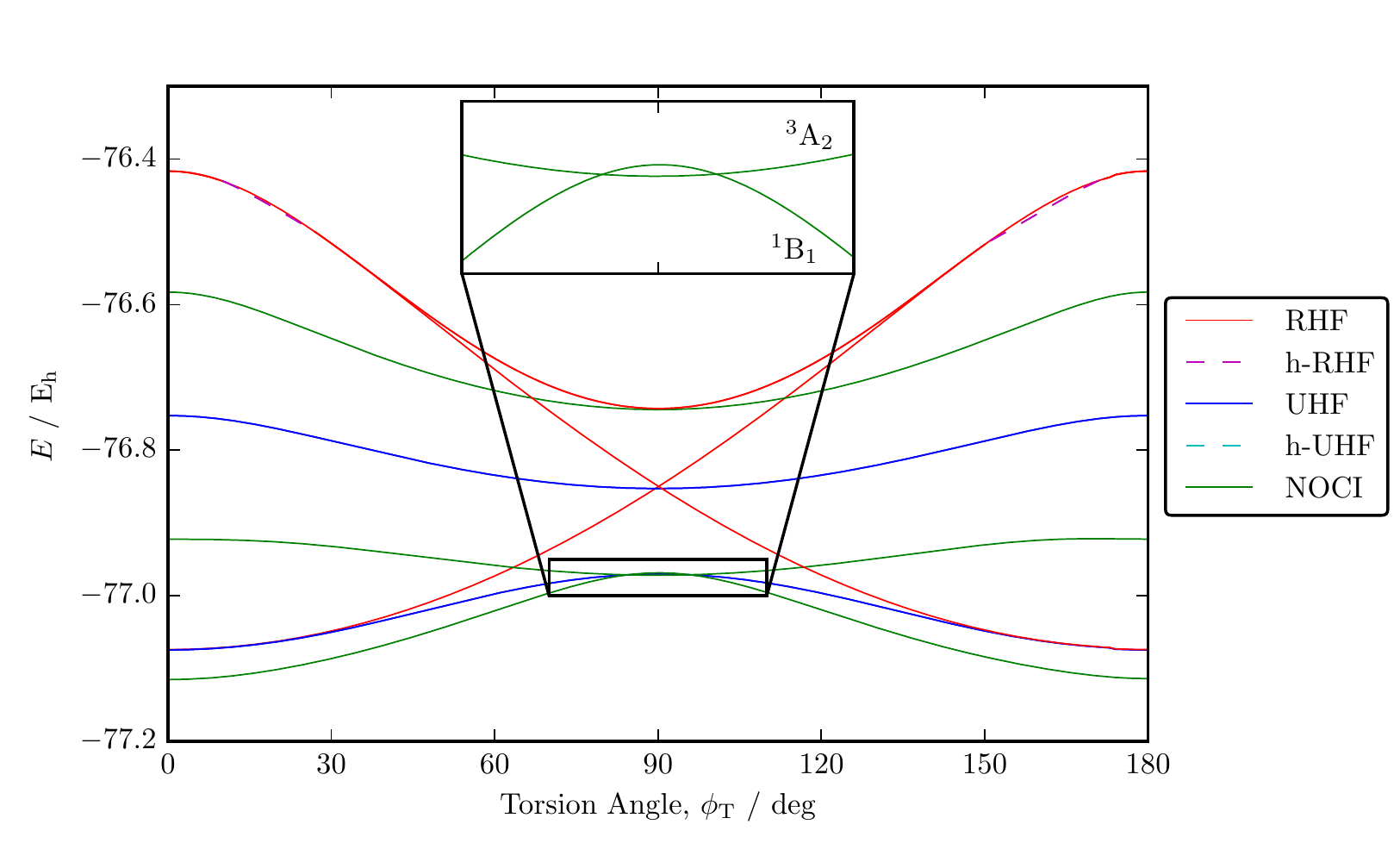}
\caption{Lowest four h-UHF (blue/cyan) and four h-RHF (red/magenta) states located using two electrons in the $\uppi$ and $\uppi^*$ orbitals of ethene (STO-3G) with a frozen core and virtual approximation.
At this bond length, the h-UHF states have real orbital coefficients (blue) across all torsion angles $\phi_{\mathrm{T}}$.
The \ce{C-H} and \ce{C-C} bond lengths and angles are fixed at their optimal values and the virtual orbitals relaxed as $\phi_{\mathrm{T}}$ varies between $0^{\circ}$ and $180^{\circ}$.
Using these solutions as a basis for NOCI recovers the $\mathrm{^{1}B_{1}}$, $\mathrm{^{3}A_{2}}$ and $\mathrm{^{1}B_{2}}$ states (green), predicting the crossing of the lowest energy singlet and triplet surfaces in agreement with the exact FCIQMC results for the STO-3G basis set.}
\label{fig:C2H4_r130_NOCI}
\end{figure*}

We next use these states as a basis for NOCI to investigate the multireference energy levels as $\phi_{\mathrm{T}}$ varies, whilst retaining the STO-3G optimised bond lengths and angles. 
Increasing $\phi_{\mathrm{T}}$ from $0^{\circ}$  to $90^{\circ}$ causes the energies of the bonding $\uppi$ and anti-bonding $\uppi^*$ orbitals to converge, forming a degenerate $\mathrm{e}$ molecular orbital pair.
Consequently, the perpendicular $\mathcal{D}_{\mathrm{2d}}$ structure consists of the nearly degenerate $\mathrm{^{1}B_{1}}$ and $\mathrm{^{3}A_{2}}$ states accompanied by the higher energy $\mathrm{^{1}B_2}$ and $\mathrm{^{1}A_1}$ states.
These correlate respectively to the $\mathrm{^{1}A_{g}}$, $\mathrm{^{3}B_{1u}}$, $\mathrm{^{1}B_{1u}}$, and $\mathrm{^{1}A_{g}}$ states at the planar $\mathcal{D}_{\mathrm{2h}}$ geometry.

The correct ordering of the $\mathrm{^{1}B_{1}}$ and $\mathrm{^{3}A_{2}}$ states has long been a subject of particular interest, with general valence bond\cite{Voter1985} and multiconfigurational SCF\cite{Brooks1979, Schmidt1987, Benassi2000, Oyedepo2010} calculations both indicating the $\mathrm{^{1}B_{1}}$ state lies below the $\mathrm{^{3}A_{2}}$ state in a rare violation of Hund's rules.\cite{Walsh1953, Merer1969}
To capture the triplet states using NOCI, spin contaminated h-UHF solutions must be included in the basis set.
For this particular case where $n=2$, our formal mathematical treatment indicates a further four h-UHF states exist across all geometries.
Retaining the frozen core and virtual orbital approximations, these additional h-UHF solutions have real orbital coefficients for all $\phi_{\mathrm{T}}$, corresponding to the diradical states and the $\left( \uppi \right)^1 \left( \uppi^* \right)^1$ configurations.
Unfreezing the virtual orbitals and relaxing the SCF states then allows the inclusion of hyperconjugation with the C-H $\upsigma^*$ orbitals.
Using these eight solutions, the three lowest NOCI energy levels are computed as shown in Figure \ref{fig:C2H4_r130_NOCI}.
With this basis set and carbon-carbon bond length only the ionic h-RHF states become complex, however including these states is essential to prevent discontinuities in the singlet NOCI energy levels.

The NOCI results presented in Figure \ref{fig:C2H4_r130_NOCI} indicate that the $\mathrm{^{3}A_{2}}$ state lies below the  $\mathrm{^{1}B_{1}}$ state at the $90^{\circ}$ transition structure, as predicted by Hund's rules.\cite{Walsh1953, Merer1969}
However, Schmidt \textit{et al.} note that such results can arise when only the two $\uppi$ electrons are correlated\cite{Schmidt1987} --- for example in the two-configuration SCF calculations of Yamaguchi \textit{et al.}\cite{Yamaguchi1983} --- whilst the correct ordering requires correlation with the core electrons to be included.
To verify our NOCI results, we compute the exact, fully correlated energies of the $\mathrm{^{1}B_{1}}$ and $\mathrm{^{3}A_{2}}$ states within the STO-3G basis set using Full Configuration Interaction Quantum Monte--Carlo (FCIQMC)\cite{Booth2009} and obtain energies of $-76.98253(5)$ $\mathrm{E_h}$  and $-76.98833(5)$ $\mathrm{E_h}$ respectively, confirming the ordering predicted by NOCI.
Further comparison with the FCIQMC results indicates that, despite only including 8 out of $1.1 \times 10^7$ determinants from the full Hilbert space, NOCI captures 93\% and 92\% of the $\mathrm{^{1}B_{1}}$ and $\mathrm{^{3}A_{2}}$ correlation energies.

Although these NOCI energy levels suggest the $\mathrm{^{1}B_{1}}$ and $\mathrm{^{3}A_{2}}$ states do cross in the rotation of ethene, it is important to remember that this is a minimal basis set calculation ignoring any geometrical relaxation for the triplet $\mathrm{^{3}A_{2}}$ state or the non-planar structures. 
Regardless, it is reassuring to observe the qualitative accuracy of NOCI within the STO-3G basis set approximation using a minimal number of determinants and a frozen core approximation.
Furthermore, the occurrence of complex h-RHF solutions as the molecular control parameters vary highlights the important role of holomorphic Hartree--Fock theory if NOCI is to be applied over all ranges of molecular geometries and compositions.

\section{Computational details}

Calculations to locate h-RHF and h-UHF solutions were performed using a holomorphic analogue to the Geometric Direct Minimisation\cite{VanVoorhis2002} method implemented with processing from \texttt{SciPy}.\cite{SciPy}
FCIQMC energies were obtained using the \texttt{HANDE 1.1}\cite{Spencer2015} stochastic quantum chemistry package.
All one- and two-electron integrals were computed in \texttt{Q-Chem 4.3}\cite{QChem4-0} whilst all figures were plotted using \texttt{Matplotlib}\cite{Matplotlib}.

\section{Conclusions}

In this work we have highlighted the properties and behaviour of h-RHF solutions for two-electron problems. 
By formulating the h-RHF problem in the framework of algebraic geometry, the exact number of h-RHF states (counted with multiplicity) has been identified as $\frac{1}{2} \left(3^n - 1\right)$, where $n$ is the number of basis functions. 
Consequently, h-RHF states exist for all geometries and atomic charges, and always provide a continuous basis for NOCI.
Furthermore, this expression provides an upper bound on the number of real RHF states, rigorously proving the result obtained by Stanton.\cite{Stanton1968}
We believe that algebraic geometry will also yield a generalised result for unrestricted or multiple electron systems, although it may be challenging to obtain a closed formula for these cases.

Through an in-depth study of \ce{HZ}, \ce{HHeH^2+}, \ce{HHeH} and ethene we have demonstrated the behaviour of h-RHF states as molecular geometry or atomic charges are changed.
For \ce{HZ} and ethene, the presence of molecular symmetry determines whether real RHF states coalesce at a triply degenerate confluence or a doubly degenerate pair annihilation point, although complex holomorphic states emerge in both cases.
By applying the generalised framework of catastrophe theory, we have illustrated the influence of molecular control parameters including geometry and atomic compositions on the type of these coalescence points.
Moreover, we have identified dormant h-RHF states with complex orbital coefficients across all geometries.
These states are not observed in standard Hartree--Fock but can be smoothly evolved into real RHF states by changing geometry or atomic charges and represent analytic continuations of the corresponding real RHF states.

Further investigating the h-RHF states of \ce{HHeH^2+} and \ce{HHeH} in STO-3G demonstrates the isomorphism between systems with two electrons and systems with two electron holes.
Exploiting this isomorphism allows the number of h-RHF states to be identified for both types of system.
Comparing to the upper bound of $\frac{1}{2} \times 4^n$ real RHF states obtained by Fukutome\cite{Fukutome1971} indicates that the number of h-RHF states provides a new reduced upper bound for systems with two electron holes.

Finally, by considering the $\uppi$ electrons in ethene as a two-electron problem, we have used the four h-RHF states and four h-UHF as a basis for NOCI to identify a crossing of the lowest energy singlet and triplet states at a torsion angle of $90^{\circ}$. 
Comparing with the exact STO-3G energies computed using FCIQMC then verifies this result within the basis set approximation, demonstrating the potential of combining holomorphic Hartree--Fock theory and NOCI.

Ultimately, the understanding on the nature of h-RHF solutions developed in this study provides a stronger platform for exploiting holomorphic states as a basis for NOCI, whilst also providing insight into the nature of Hartree--Fock states in general.

H.G.A.B. thanks the Cambridge and Commonwealth Trust for a Vice--Chancellor's Award Scholarship and A.J.W.T. thanks the Royal Society for a University Research Fellowship (UF110161). 
We also acknowledge Dr. James Farrell for insightful discussions and assistance.

\bibliography{EnumeratingHoloHF.bib}

\end{document}

%% file: figures/TikZ_circle_projection.tex
\begin{tikzpicture}[ scale=0.65]
\draw[->,thick] (-5,0)--(5,0) node[above]{$c_2$};
\draw[->,thick] (0,-5)--(0,5) node[right]{$c_1$};


\draw[thick] (0,0) circle  (3); 
\draw[thick,blue] (3,-5) -- (3,5) ; 
\draw[thick,blue] (-3,-5) -- (-3,5) ; 

\draw[fill=black] (1.928362829,2.29813329)  circle  (0.2); 
\draw[fill=black] (1.928362829,-2.29813329)  circle  (0.2); 
\draw[fill=black] (3,0)  circle  (0.2); 
\draw[fill=black] (0,3)  circle  (0.2); 

\draw[thick] (-1.928362829,2.29813329)  circle  (0.2); 
\draw[thick] (-1.928362829,-2.29813329)  circle  (0.2); 
\draw[thick] (-3,0)  circle  (0.2); 
\draw[thick] (0,-3)  circle  (0.2); 

\draw[fill=black] (3,-3.575260778) circle  (0.2) {};
\draw[fill=black] (3,3.575260778) circle  (0.2) {};
\draw[fill=black] (3,0) circle  (0.2) {};

\draw[dashed] (0,0) -- (3,-3.575260778);
\draw[dashed] (0,0) -- (3,3.575260778);

\draw (3,4.8) node[right]{$c_2 = 1$};
\draw (-3,4.8) node[right]{$c_2 = -1$};
\end{tikzpicture}

%% file: figures/OrbitalPlots.tex



\begin{tikzpicture}
\node[inner sep=0pt] (Surface) at (0,0)
    {\includegraphics[width=1.4\figurewidth, keepaspectratio, trim={4cm 2cm 0cm 0cm}, clip]{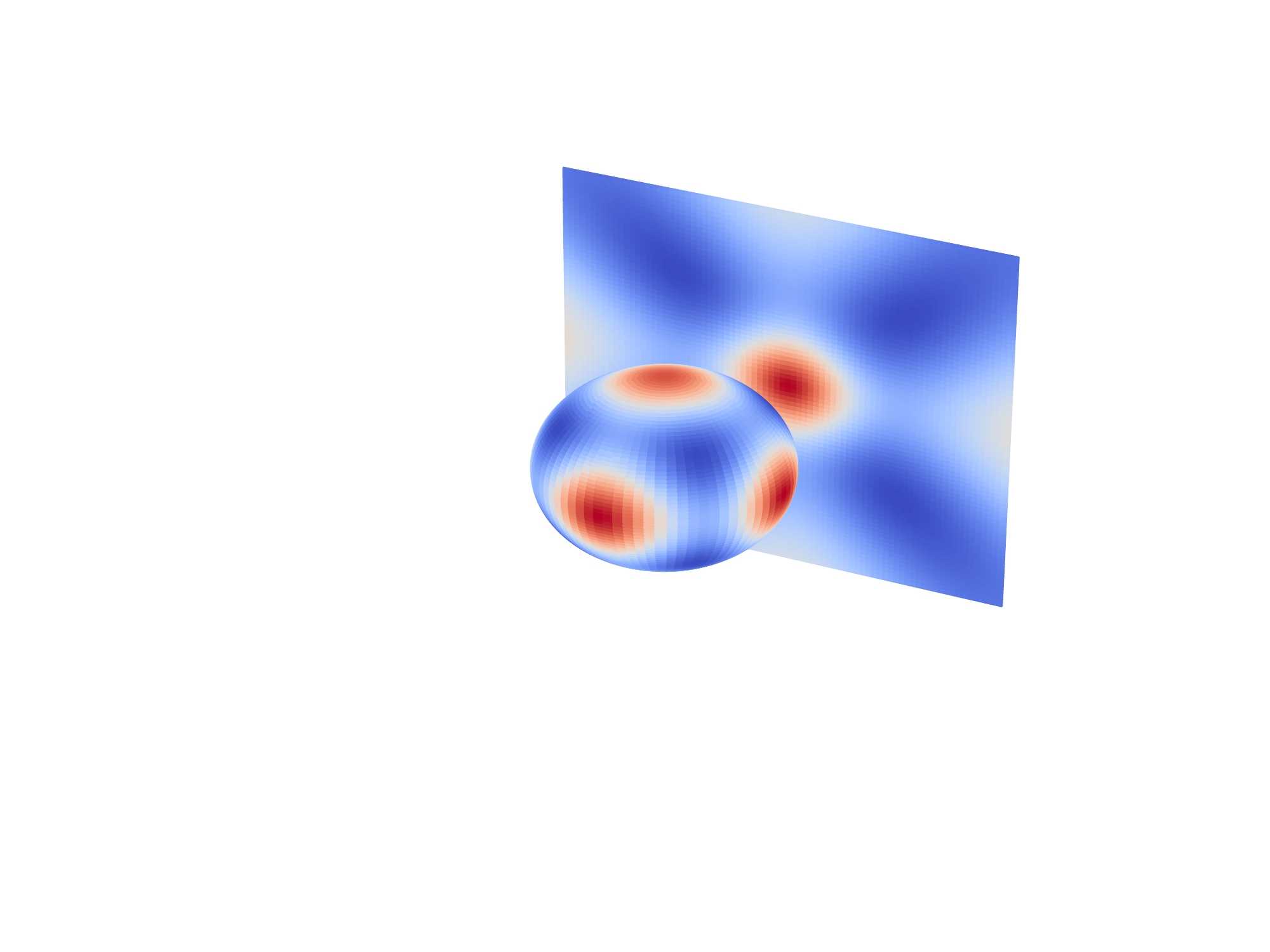}};
 
\node[inner sep=0pt] (Orb1) at (5.5,3)
    {\ovalbox{\includegraphics[width=0.3\figurewidth, height=0.1\figurewidth, trim={2cm 8cm 2cm 8cm}, clip]{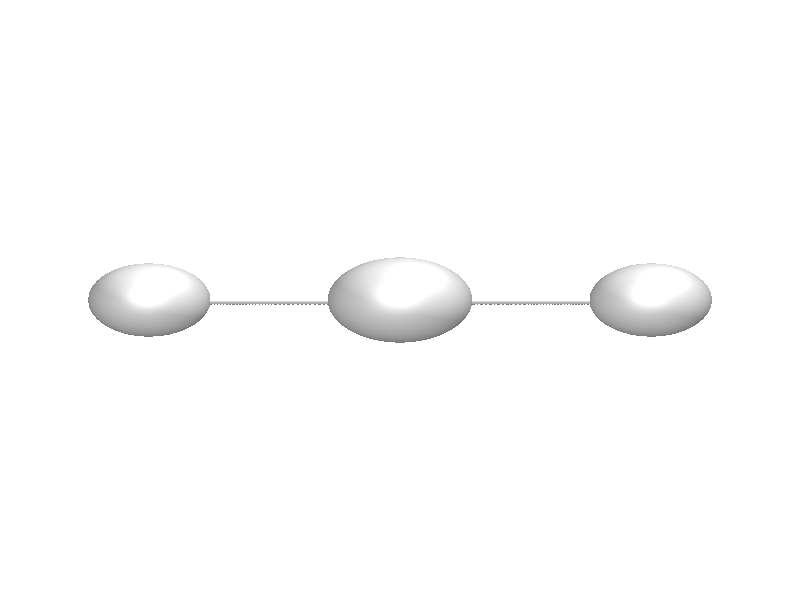}}};
\draw[fill=black] (1.9,1.3) circle (.075);  
\draw[-, thick] (1.9,1.3) -- (Orb1.west);

\node[inner sep=0pt] (Orb2) at (0.5,4)
    {\ovalbox{\includegraphics[width=0.3\figurewidth, height=0.1\figurewidth, trim={2cm 8cm 2cm 8cm}, clip]{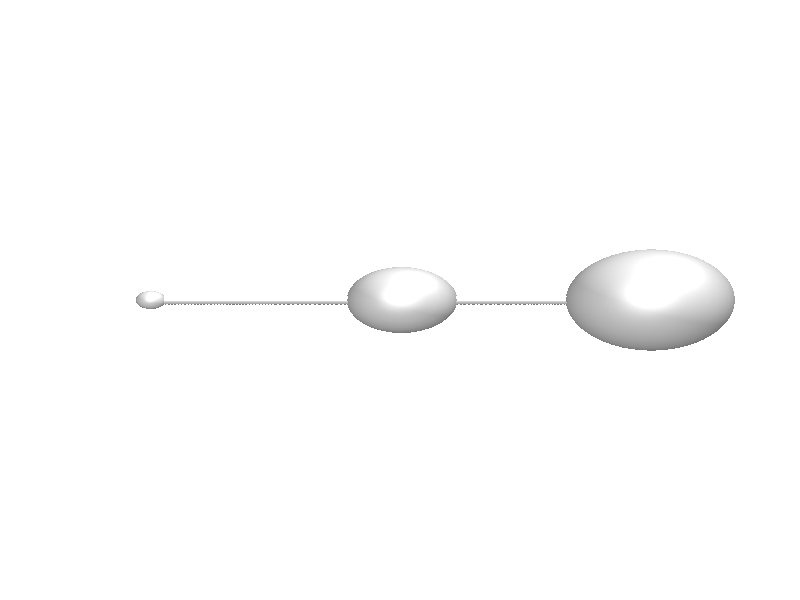}}};
\draw[fill=black] (0.45,1.55) circle (.075);  
\draw[-,thick] (0.45,1.55) -- (Orb2.south);

\node[inner sep=0pt] (Orb3) at (-5.8,4)
    {\ovalbox{\includegraphics[width=0.3\figurewidth, height=0.1\figurewidth, trim={2cm 8cm 2cm 8cm}, clip]{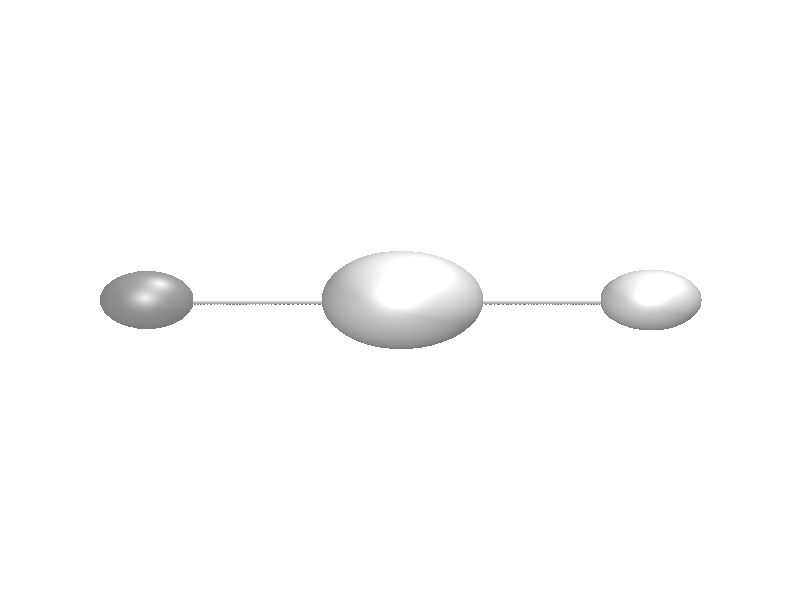}}}; 
\draw[fill=black] (-1,1.8) circle (.075);  
\draw[-,thick] (-1,1.8) -- (Orb3.east);
  
 \node[inner sep=0pt] (Orb4) at (-5.8,2)
    {\ovalbox{\includegraphics[width=0.3\figurewidth, height=0.1\figurewidth, trim={2cm 8cm 2cm 8cm}, clip]{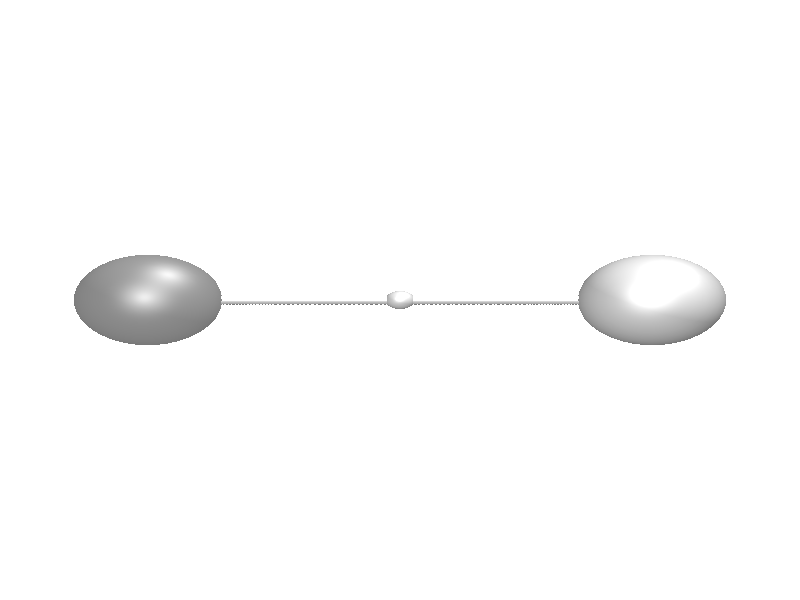}}};
\draw[fill=black] (-1.05,0.75) circle (.075);  
\draw[-, thick] (-1.05,0.75) -- (Orb4.east);

\node[inner sep=0pt] (Orb5) at (-5.8,0)
    {\ovalbox{\includegraphics[width=0.3\figurewidth, height=0.1\figurewidth, trim={2cm 8cm 2cm 8cm}, clip]{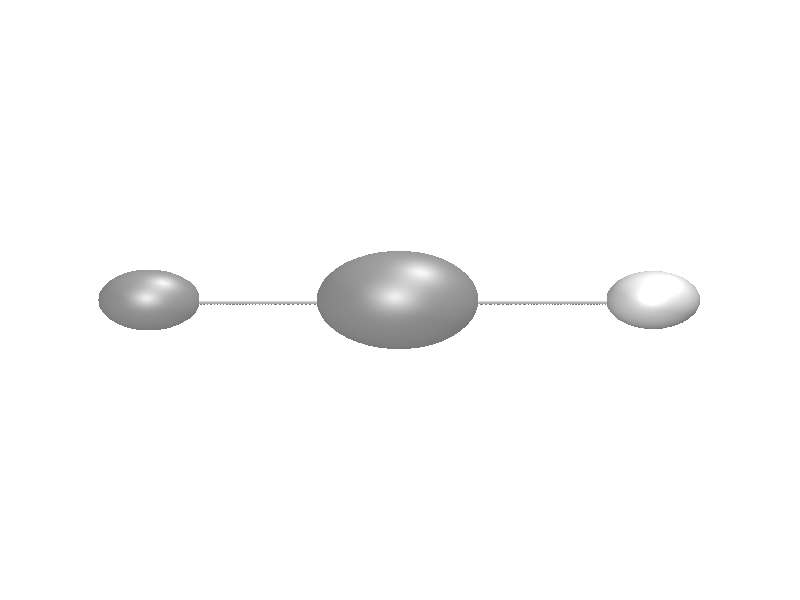}}};
\draw[-, thick] (-3,-0.2) -- (Orb5.east);

\node[inner sep=0pt] (Orb6) at (0.5,-3.25)
    {\ovalbox{\includegraphics[width=0.3\figurewidth, height=0.1\figurewidth, trim={2cm 8cm 2cm 8cm}, clip]{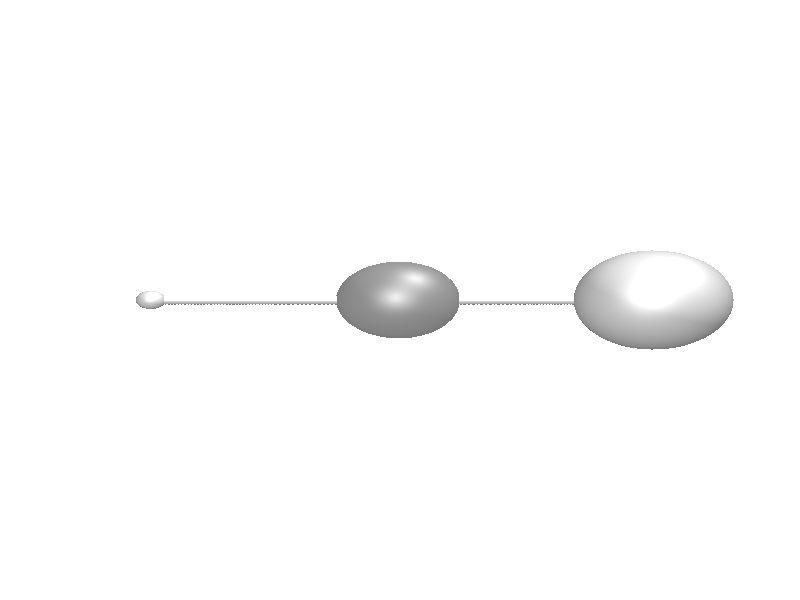}}};    
\draw[-, thick] (0.45,-1.15) -- (Orb6.north);

\node[inner sep=0pt] (Orb7) at (4.5,-4)
    {\ovalbox{\includegraphics[width=0.3\figurewidth, height=0.1\figurewidth, trim={2cm 8cm 2cm 8cm}, clip]{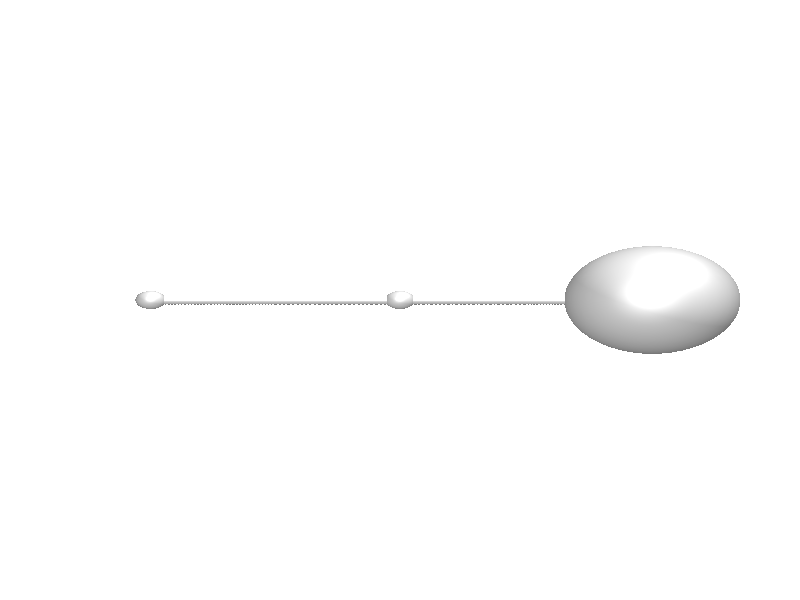}}}; 
\draw[fill=black] (0.4,0.35) circle (.075); 
\draw[-, thick] (0.4,0.35) -- (Orb7.west);

\node[inner sep=0pt] (Orb8) at (5.5,-2)
    {\ovalbox{\includegraphics[width=0.3\figurewidth, height=0.1\figurewidth, trim={2cm 8cm 2cm 8cm}, clip]{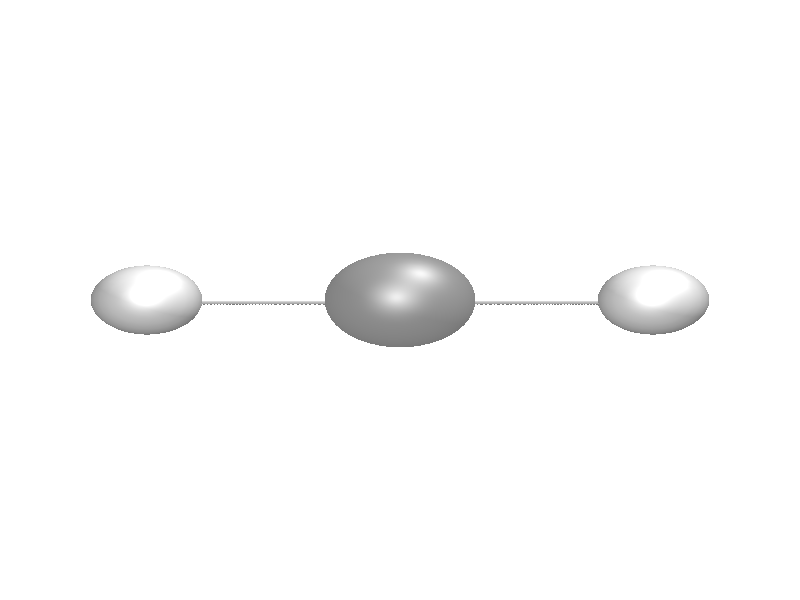}}};
\draw[fill=black] (1.85,-1.15) circle (.075);   
\draw[-, thick] (1.85,-1.15) -- (Orb8.west);

\node[inner sep=0pt] (Orb9) at (5.5,0.5)
    {\ovalbox{\includegraphics[width=0.3\figurewidth, height=0.1\figurewidth, trim={2cm 8cm 2cm 8cm}, clip]{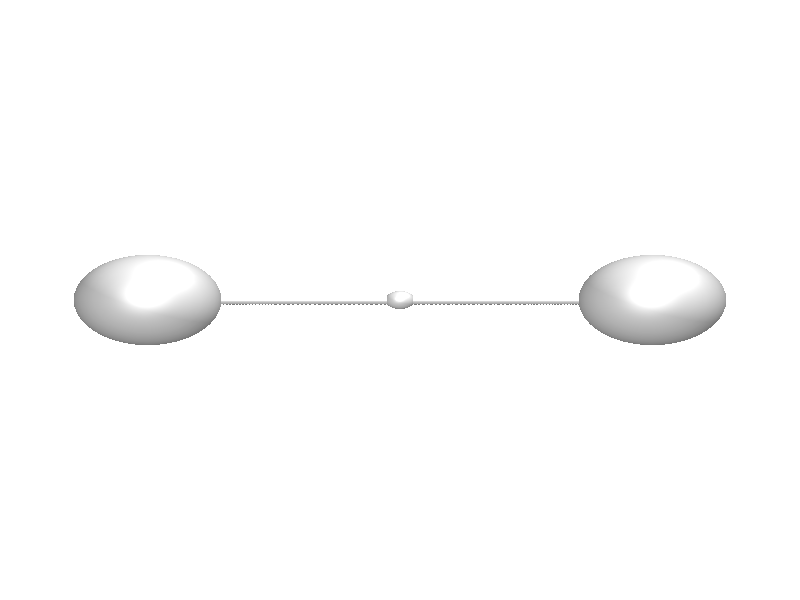}}};   
\draw[fill=black] (1.875,0.0) circle (.075);      
\draw[-, thick] (1.875,0.0) -- (Orb9.west);

\draw[->,thick] (-7,-3.5) -- (-7,-1.5) node[right] {$\ c_2$};
\draw[->,thick] (-7,-3.5) -- (-5,-4.1) node[right] {$\ c_1$};
\draw[->,thick] (-7,-3.5) -- (-5.25,-2.5) node[right] {$\ c_3$};

\end{tikzpicture}


%% file: figures/LipsBifurcation.tex
\begin{tikzpicture}
\node[inner sep=0pt] (Surface) at (0.27,1.55)
    {\includegraphics[scale=0.8, angle=0, keepaspectratio, trim={0.0cm 1.cm 7.5cm 0cm}, clip]{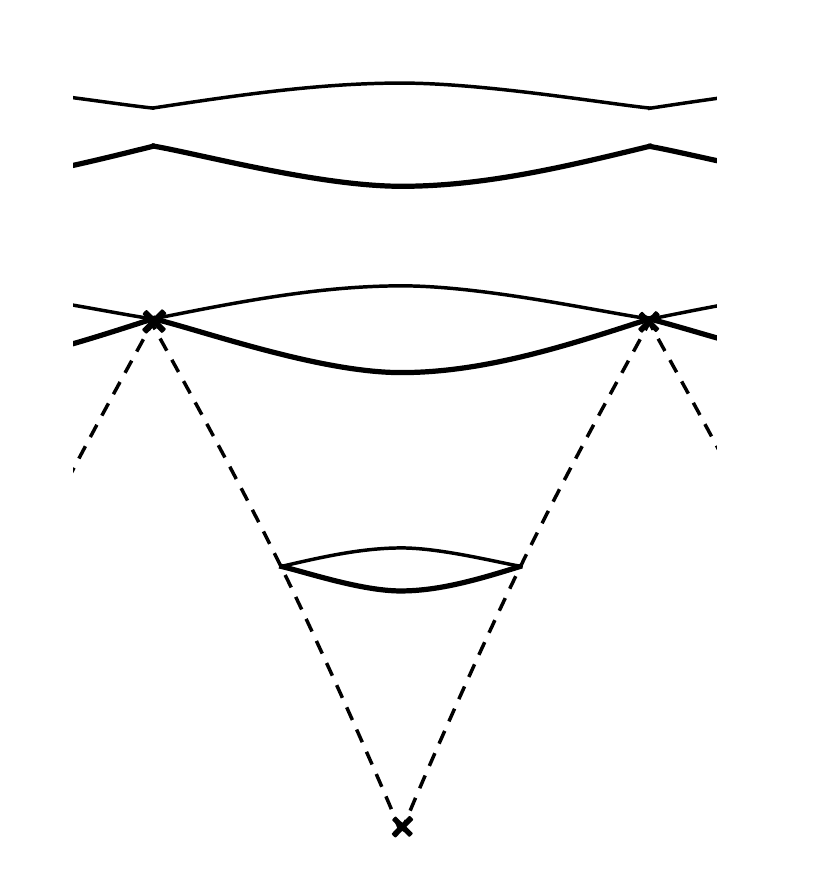}};
    
\draw[->,thick] (-3.5,-1.8) -- (-1.9,-1.8) node[above] {$\phi_{\mathrm{T}}$}; 
\draw[->,thick] (-3.5,-1.8) -- (-3.5,-0.3) node[right] {$\ R_{\mathrm{CC}}$}; 
\draw[->,thick] (-3.5,-1.8) -- (-2.7,-1.0) node[right] {$Q_{\mathrm{X}}$};

\draw[dotted,thick] (2.125,4.9) node[above] {$^{\phantom{\circ}}180^{\circ}$} -- (2.125,1.4) ;
\draw[dotted,thick] (-1.9,4.9) node[above] {$^{\phantom{\circ}}0^{\circ}$} -- (-1.9,1.4) ; 

\draw[->,thick] (1,-2) node[right] {$\mathrm{A}^{+}_{3}$} -- (0.3,-1.7);
\draw[->,thick] (3.0,3.25) node[right] {$\mathrm{A}^{-}_{3}$} -- (2.28,2.7);

\node at (0.24,0) {$\mathrm{A}^{\phantom{+}}_{2}$};
\node at (0.24,1) {$\mathrm{A}^{\phantom{+}}_{2}$};

\node at (0.24,3.95) {$\mathrm{A}^{\phantom{+}}_{2}$};
\node at (0.24,4.8) {$\mathrm{A}^{\phantom{+}}_{2}$};

\node at (1.5, 0.5) {$\mathrm{A}_{3}$};
\node at (-1.3, 0.5) {$\mathrm{A}_{3}$};

\node[inner sep=0pt] (Surface) at (8.95,1.55)
    {\includegraphics[scale=0.8, angle=0, keepaspectratio, trim={6.5cm 1.cm 1.2cm 0.0cm}, clip]{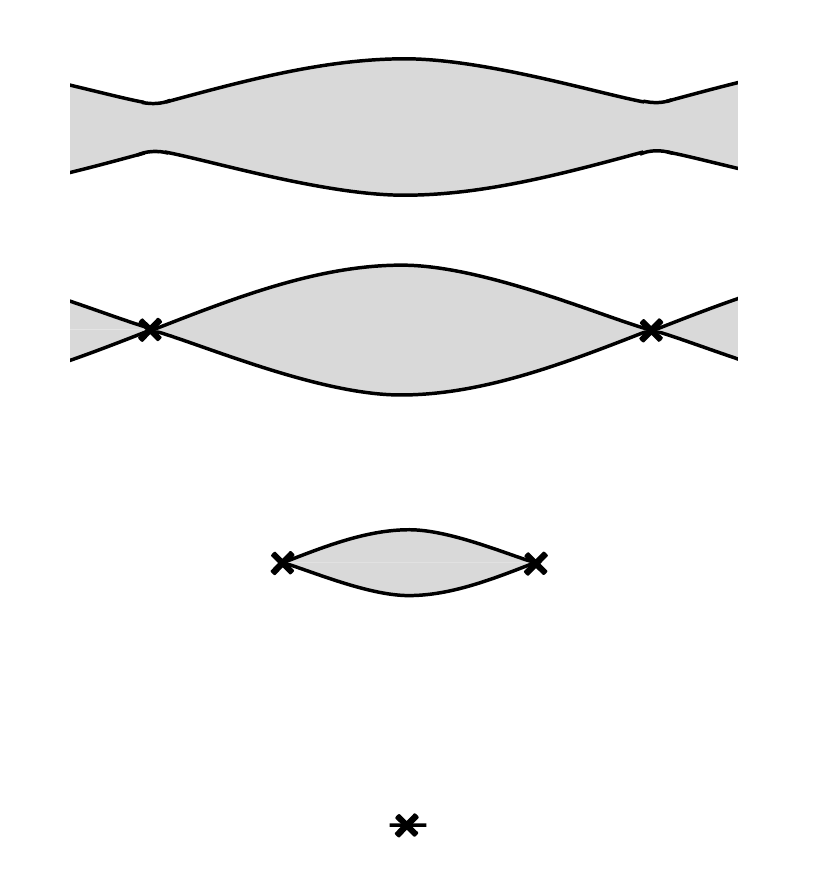}};
    
\draw[->,thick] (4.4,3.5) -- (5.6,3.5) node[above] { {\footnotesize $\phi_{\mathrm{T}}$} };
\draw[->,thick] (4.4,3.5) -- (4.4,4.7) node[right] { {\footnotesize $\ Q_{\mathrm{X}}$} };

\draw[->,thick] (4.4,1.7) -- (5.6,1.7) node[above] { {\footnotesize $\phi_{\mathrm{T}}$} };
\draw[->,thick] (4.4,1.7) -- (4.4,2.9) node[right] { {\footnotesize $\ Q_{\mathrm{X}}$} };

\draw[->,thick] (4.4,-0.1) -- (5.6,-0.1) node[above] { {\footnotesize $\phi_{\mathrm{T}}$} };
\draw[->,thick] (4.4,-0.1) -- (4.4,1.1) node[right] { {\footnotesize $\ Q_{\mathrm{X}}$} };

\draw[->,thick] (4.4,-1.9) -- (5.6,-1.9) node[above] { {\footnotesize $\phi_{\mathrm{T}}$} };
\draw[->,thick] (4.4,-1.9) -- (4.4,-0.7) node[right] { {\footnotesize $\ Q_{\mathrm{X}}$} };

\draw[dotted,thick] (10.9,4.9) node[above] {$^{\phantom{\circ}}180^{\circ}$} -- (10.9,3.5) ;
\draw[dotted,thick] (6.825,4.9) node[above] {$^{\phantom{\circ}}0^{\circ}$} -- (6.825,3.5) ; 
\draw[dotted,thick] (10.9,2.85) -- (10.9,1.8);
\draw[dotted,thick] (6.825,2.85) -- (6.825,1.8);

\node at (8.3,-1.55) {$\mathrm{A}^{+}_{3}$};

\node at (10.4, 0.5) {$\mathrm{A}_{3}$};
\node at (7.5, 0.5) {$\mathrm{A}_{3}$}; 

\node at (9.0,1.15) {$\mathrm{A}^{\phantom{+}}_{2}$};
\node at (9.0,-0.05) {$\mathrm{A}^{\phantom{+}}_{2}$};

\node at (9.0,3.9) {$\mathrm{A}^{\phantom{+}}_{2}$};
\node at (9.0,4.95) {$\mathrm{A}^{\phantom{+}}_{2}$};

\draw[->,thick] (11.5,1.8) node[below] {$^{\phantom{-}}_{\phantom{3}}\mathrm{A}^{-}_{3}$} -- (11.0,2.3);


\end{tikzpicture}